\documentclass[pra,onecolumn,nofootinbib,floatfix,10pt]{revtex4-2}
\usepackage{bm}
\usepackage{braket}
\usepackage{amsfonts}
\usepackage{tabularx}
\usepackage{graphicx}
\usepackage{color,soul}
\usepackage{siunitx}
\usepackage{amsmath}
\usepackage{amssymb}
\usepackage{multirow}
\usepackage{wasysym}
\usepackage{booktabs}
\usepackage{color,soul}
\usepackage{physics}
\usepackage{dsfont}
\usepackage{float}
\usepackage[english]{babel}
\usepackage{blindtext}
\usepackage[english,nomargin,inline,marginclue,draft]{fixme}
\pdfpageheight\paperheight
\pdfpagewidth\paperwidth

\usepackage[colorlinks,linkcolor=blue,anchorcolor=blue,citecolor=blue,urlcolor=blue]{hyperref}

\fxusetheme{colorsig}
\FXRegisterAuthor{cg}{acg}{CG}  
\FXRegisterAuthor{th}{ath}{\color{blue}TH}  
\FXRegisterAuthor{ib}{aib}{\color{red}IB} 
\FXRegisterAuthor{sh}{ash}{\color{cyan}SH} 
\FXRegisterAuthor{db}{adb}{\color{green}DB} % now I can ..
\FXRegisterAuthor{ps}{aps}{PS}
\makeatletter
\renewcommand*\FXLayoutInline[3]{%
	{\@fxuseface{inline}\ignorespaces{\color{fx#1}[#3: #2]}}}
\makeatother

\long\def\symbolfootnote[#1]#2{\begingroup%
	\def\thefootnote{\fnsymbol{footnote}}\footnotetext[#1]{#2}\endgroup}

\def\nobreakbefore{%
	\relax\ifvmode\else
	\ifhmode
	\ifdim\lastskip > 0pt\relax
	\unskip\nobreakspace
	\else % added to put a ~if no space was typed. (Unclear why it sometimes worked before )
	\nobreakspace
	\fi
	\fi
	\fi
}
\let\oldcite\cite
\renewcommand\cite{\nobreakbefore\oldcite}

\begin{document}
	\title{Non-Hermitian physics in the dissipative many-body system of Rydberg atoms}
	
	\author{Ya-Jun Wang$^{1,2,\textcolor{blue}{\dagger}}$}
	\author{Jun Zhang$^{1,2,\textcolor{blue}{\dagger}}$}
	\author{Dong-Sheng Ding$^{1,2,\textcolor{blue}{\star}}$}
	
	\affiliation{$^1$Key Laboratory of Quantum Information, University of Science and Technology of China, Hefei, Anhui 230026, China.}
	\affiliation{$^2$Synergetic Innovation Center of Quantum Information and Quantum Physics, University of Science and Technology of China, Hefei, Anhui 230026, China.}
	
	\date{\today}
	
	\symbolfootnote[1]{dds@ustc.edu.cn}
	
	\maketitle
	\section{Abstract}
	Non-Hermitian physics exhibits unique physical properties beyond those of traditional Hermitian systems, such as symmetry breaking, the emergence of exceptional points, topological phase transitions, and more. These phenomena have been extensively studied across various platforms, including quantum optics, cold atom systems, superconducting circuits, and condensed matter physics. Rydberg atoms, with their long-range interactions and flexible controllability, provide a promising platform for the experimental realization of non-Hermitian physics. This review primarily summarizes the key experimental and theoretical achievements in the field of non-Hermitian physics within Rydberg atomic systems in recent years. It outlines the fundamental construction of non-Hermitian Hamiltonians, reveals the effective dissipation mechanisms induced by Rydberg atomic interactions, and discusses their impact on spectral properties and symmetry breaking. These studies not only deepen the understanding of quantum phase transitions in non-Hermitian many-body systems but also highlight the unique value of Rydberg atomic platforms in realizing and controlling topological states.
	
\section{Introduction}
In conventional quantum mechanics, Hamiltonians are typically Hermitian, ensuring real-valued energy eigenvalues. However, non-Hermitian Hamiltonians allow for complex eigenvalues \cite{yokomizo2019non,kawabata2019symmetry,ashida2020non,bergholtz2021exceptional}, corresponding to non-conservative dynamics with gain or dissipation, and exhibit unique physical behaviors across a wide range of fields. For example, in condensed matter systems, the finite quasiparticle lifetime caused by interactions or disorder can introduce non-Hermitian effects~\cite{shen2018quantum,nagai2020dmft}; in the system of interaction between light and medium, absorption or radiative loss in the medium can also introduce non-Hermitian terms into the wave equation~\cite{wang2022non,li2023synergetic,chen2023weak,han2023exceptional,xiao2025non}. Importantly, Non-Hermiticity enriches the manifestations of core physical concepts such as symmetry breaking and topological phase transitions~\cite{helbig2020generalized,wang2021topological,zhang2022review,li2025realization}. In Hermitian systems, symmetry may undergo spontaneous breaking at critical points (e.g., magnetic ordering in quantum spin systems). In contrast, the introduction of dissipation and gain in non-Hermitian systems serves as an additional tuning mechanism, thereby enabling a greater diversity of symmetry breaking phenomena. For instance, non-Hermiticity can induce the spontaneous breaking of time-reversal symmetry~\cite{beasley1994time,koch2010time,kawabata2019topological,ghosh2020recent} or parity-time symmetry~\cite{mandal2021symmetry,lourencco2022non,el2018non,li2019observation}. Meanwhile, significant progress has been made in recent years in the study of topological states of matter within non-Hermitian systems~\cite{bonderson2013quasi,ren2016topological,ma2019topological}, as explored across diverse platforms including optics~\cite{barik2018topological,parto2020non}, superconducting circuits~\cite{imhof2018topolectrical,roushan2014observation}, and cold atom systems~\cite{li2020topological,yoshida2018reduction}. Although non-Hermiticity introduces fundamental differences in aspects such as spectral structure and eigenstate orthogonality, many topological phases can still be defined and realized under appropriate frameworks of symmetry and topological classification, such as non-Hermitian versions of the Su-Schrieffer-Heeger model~\cite{lieu2018topological,hao2023topological}, Chern insulators~\cite{philip2018loss,chen2018hall}, quantum spin Hall insulators~\cite{kane2005z}, topological lasers~\cite{harari2018topological,bandres2018topological}, and bulk Fermi arcs connecting exceptional points~\cite{zhou2018observation,kozii2024non,hu2025unconventional}. These emergent phenomena have profoundly expanded our understanding of the fundamental physics governing non-Hermitian and open quantum systems.

Rydberg atomic systems, leveraging their long-range interactions and high controllability~\cite{busche2017contactless,maxwell2013storage,dudin2012strongly}, have emerged as an important experimental platform for exploring various complex quantum effects~\cite{saffman2010quantum,adams2019Rydberg,browaeys2020many,ding2023ergodicity}. Through laser manipulation and tunable environmental coupling, researchers have successfully realized a series of quantum states and dynamical behaviors in Rydberg systems, such as Bose-Einstein condensation~\cite{balewski2013coupling,chien2024breaking}, quantum phase transitions~\cite{carr2013nonequilibrium, perez2018glassy,heyl2018dynamical,ding2019Phase}, topological phases~\cite{de2019observation,zhang2025observation}, and time crystals~\cite{gambetta2019,wu2024dissipative,liu2024higher,liu2025bifurcation}. More importantly, the inherent dissipative pathways and many-body interactions in Rydberg ensembles provide a natural avenue for engineering tunable non-Hermitian Hamiltonians, thereby extending the scope of research from traditional Hermitian systems to non-Hermitian quantum physics. 

In non-Hermitian Rydberg atom platforms, symmetry breaking can occur not only in static phases but can also emerge dynamically during time evolution, giving rise to phenomena such as spontaneous charge-conjugate parity or particle-hole symmetry breaking~\cite{zhang2025exceptional,wang2026quantum}, non-Hermitian-induced complex dynamical hysteresis trajectories~\cite{zhang2025exceptional,xie2025chiral}, and exceptional points and their applications in enhanced sensing~\cite{wang2026quantum,vsumarac2026controlling,liang2026exceptional}. Furthermore, Rydberg atomic systems, with their highly tunable array structures, dipole exchange interactions, and close connection to quantum information processing technologies, provide a highly promising platform for realizing, tuning, and detecting topological states and non-Hermitian quantum phase transitions. Through precise design of atomic arrays~\cite{zhang2025observation1}, synthetic dimensions~\cite{kanungo2022realizing,lu2024probing,lu2024wave}, and dissipative couplings~\cite{zhang2025observation}, quantum simulators with nontrivial topological or non-Hermitian properties can be constructed in these systems, thus opening avenues for exploring frontier topics.

The integration of non-Hermitian physics with many-body systems of Rydberg atoms provides a powerful platform for exploring the non-equilibrium, non-Hermitian, and topological properties of quantum many-body physics. In this review, we primarily summarize the key experimental and theoretical advances related to Rydberg atomic systems and non-Hermitian physics. Starting from the fundamental construction of non-Hermitian Hamiltonians, we elucidate the effective dissipation mechanisms induced by Rydberg atomic interactions and analyze their impact on spectral properties and symmetry breaking. Subsequently, the emergence, dynamical behavior, and potential applications of exceptional points in quantum sensing are discussed. In addition, we outline the implementations of both topological and non-Hermitian quantum phase transitions in Rydberg atom platforms (including ensembles, synthetic dimensions, and arrays), along with a topological dissipative structure based on the Su-Schrieffer-Heeger model. These studies not only deepen the understanding of quantum phase transitions in many-body systems but also highlight the unique value of the Rydberg platform for realizing and controlling topological states. 

The structure of this review is as follows. Section \ref{sec1} introduces the fundamental theoretical framework of non-Hermitian physics, including non-Hermitian Hamiltonians, symmetry, topological properties, the Lindblad master equation, Liouvillian exceptional points, and the development roadmap of non-Hermitian physics in Rydberg atomic systems. Section \ref{sec2} summarizes recent experimental advances in non-Hermitian phenomena in Rydberg atomic systems, covering hysteresis loops, exceptional points, chiral state switching, and related topics. Section \ref{sec3} discusses non-Hermitian enhanced sensing based on Rydberg atomic platforms. Section \ref{sec4} presents the realization of topological states and non-Hermitian quantum phase transitions in Rydberg atomic systems, including spectral topology, the dissipative Su-Schrieffer-Heeger model, synthetic dimensions, and parity-time symmetry breaking. Section \ref{sec5} concludes the review and provides an outlook on future research directions.

\section{Non-Hermitian physics}\label{sec1}
\subsection{The basic theory of non-Hermitian systems}
In quantum physics, a defining feature of non-Hermitian systems is that their eigenstates no longer satisfy orthogonality. Instead, they are characterized by a biorthonormal system, composed of left and right eigenvectors. The study of these systems is crucial for understanding a wide range of physical phenomena. These properties not only challenge the framework of traditional quantum mechanics but also provide valuable perspectives for investigating the properties of open quantum systems, dissipative phase transitions, and other key concepts. Due to $H\neq H^\dagger$ in non-Hermitian systems, it is necessary to introduce a biorthogonal basis~\cite{brody2013biorthogonal}. Considering a non-Hermitian Hamiltonian 
\begin{align}
	H_{\mathrm{NH}} &= H + i\Gamma \quad H^{\dagger}_{\mathrm{NH}} = H - i\Gamma, 
\end{align}
its eigenvalue equation can be written as
\begin{align}
	H_{\mathrm{NH}}|\psi_{n}^{R}\rangle &= E_{n}|\psi_{n}^{R}\rangle, \quad \langle \psi_{n}^{R}|H^{\dagger}_{\mathrm{NH}} = E_{n}^{*}\langle \psi_{n}^{R}|, \\
	H^{\dagger}_{\mathrm{NH}}|\psi_{n}^{L}\rangle &= E_{n}^{*}|\psi_{n}^{L}\rangle, \quad \langle \psi_{n}^{L}|H_{\mathrm{NH}} = E_{n}\langle \psi_{n}^{L}|.
\end{align}
$E_n$ and $E_n^*$ are the corresponding eigenvalues, where $n$ is the index used to distinguish the eigenvalues. If the eigenvalues of the Hamiltonian are non-degenerate, then although the eigenspace of the Hamiltonian is complete, the eigenstates do not satisfy orthonormality.
For two right eigenstates ${ \ket{\psi_m^R} }$ and ${ \ket{\psi_n^R} }$, based on $2H = H_{\text{NH}} + H_{\text{NH}}^\dagger$ and $2i\Gamma = H_{\text{NH}} - H_{\text{NH}}^\dagger$, we obtain
\begin{align}
	\bra{\psi_m^R} 2H \ket{\psi_n^R} = \bra{\psi_m^R} (H_{\text{NH}} + H_{\text{NH}}^\dagger) \ket{\psi_n^R} = (E_n + E_m^*) \braket{\psi_m^R | \psi_n^R},
\end{align}
\begin{align}
	\bra{\psi_m^R} 2i\Gamma \ket{\psi_n^R} = \bra{\psi_m^R} (H_{\text{NH}} - H_{\text{NH}}^\dagger) \ket{\psi_n^R} = (E_n - E_m^*) \braket{\psi_m^R | \psi_n^R},
\end{align}
\begin{align}
	\braket{\psi_m^R | \psi_n^R} = 2 \frac{\bra{\psi_m^R} H \ket{\psi_n^R}}{E_m^* + E_n} = 2i \frac{\bra{\psi_m^R} \Gamma \ket{\psi_n^R}}{E_n - E_m^*} \neq 0.
\end{align}
This indicates that different right eigenstates are non-orthogonal. However, by introducing the biorthogonal basis 
\begin{align}
	\bra{\psi_m^L} H_{\text{NH}} \ket{\psi_n^R} = E_n \braket{\psi_m^L | \psi_n^R} = E_m \braket{\psi_m^L | \psi_n^R},
\end{align}
we find that if $E_m \neq E_n$, then $\braket{\psi_m^L | \psi_n^R}$ must equal zero. For a general non-Hermitian Hamiltonian, its eigenvalues are typically complex unless specific symmetries are present. If the left and right eigenstates are linearly independent, they can be biorthogonalized
\begin{align}
	\ket{\tilde{\psi}_n^R} = \frac{\ket{\psi_n^R}}{\sqrt{\braket{\psi_n^L|\psi_n^R}}}, \quad \ket{\tilde{\psi}_n^L} = \frac{\ket{\psi_n^L}}{\sqrt{\braket{\psi_n^L|\psi_n^R}}}.
\end{align}
The biorthogonalized eigenstates satisfy the following orthonormality and completeness relations
\begin{align}
	\braket{\tilde{\psi}_m^L | \tilde{\psi}_n^R} = \delta_{mn}, \quad \sum_n \ket{\tilde{\psi}_n^R} \bra{\tilde{\psi}_n^L} = \mathbb{I}.
\end{align}

At present, the physical realization of non-Hermitian systems primarily relies on two mechanisms: non-reciprocal coupling and gain or loss. In lattice models, non-reciprocal coupling manifests as the non-reciprocity of hopping amplitudes between adjacent sites, meaning that the coupling strength from site $j$ to site $j+1$, denoted as $t_{j\rm{L}}$, is not equal to the reverse coupling $t_{j\rm{R}}$~\cite{jiang2020topological}, leading to  the directional transport of energy or particle flow. On the other hand, the gain and loss mechanism corresponds to introducing a complex on-site potential at each lattice site, where the imaginary part represents gain (positive imaginary part) or loss (negative imaginary part)~\cite{zeng2016non}. This is often achieved in practical systems through optical pumping, radiative decay, or coupling to external environments.

Precise control of non-reciprocal coupling typically requires specially designed temporal modulation or nonlinear components, making it relatively challenging to implement. In contrast, gain and loss are more commonly encountered in open systems such as photonic~\cite{shen2016experimental,feng2017non}, acoustic~\cite{ghatak2020observation,fruchart2021non}, and cold atom platforms~\cite{li2019observation,gou2020tunable}. Furthermore, by spatially engineering the distribution of dissipation, gain‑loss structures can be effectively constructed~\cite{ashida2020non,li2024loss}. Therefore, studying non‑Hermitian systems from the perspective of gain and loss not only offers greater experimental feasibility but also provides a practical framework for exploring Non-Hermitian physics in a wide range of open systems.
\subsection{Symmetry in non-Hermitian systems}
Symmetry and its breaking are central to understanding phase transitions and the classification of quantum states in many-body systems~\cite{ji2020categorical,chen2023continuous}. The most fundamental symmetry is internal (non-spatial) symmetry, which does not depend on any specific spatial structure. Within Hermitian systems, the key internal symmetries are formalized by the Altland-Zirnbauer classification~\cite{chiu2013classification}: time-reversal symmetry, particle-hole symmetry~\cite{chiu2016classification}, and chiral symmetry. 

In non-Hermitian systems, symmetry continues to play a crucial role in governing topological phases. A principal distinction between Hermitian and non-Hermitian systems lies in their accessible degrees of freedom. Non-Hermitian systems can realize non-unitary operations that are forbidden in Hermitian contexts. This ability allows the energy spectrum to shift from the real axis to the complex plane, thereby expanding the parameter space needed to describe the system. As a result, the study of symmetry in non-Hermitian physics not only generalizes Hermitian topological concepts but also predicts the emergence of topological phases. Indeed, non-Hermiticity fundamentally splits and unifies symmetry. To illustrate symmetry splitting, we take particle-hole symmetry as an example. In a Hermitian system, particle-hole symmetry is defined as~\cite{kawabata2019symmetry}
\begin{align}\label{9}
	\mathcal{C} H^* \mathcal{C}^{-1}=-H,
\end{align}
where $\mathcal{C}$ is a unitary matrix. For non-Hermitian systems, however, $H^* \neq H^T$ (where $ H^*$ denotes the complex conjugate of the Hamiltonian, and $H^T$ denotes the transpose of the Hamiltonian); therefore, the above definition is not equivalent to the one based on transposition~\cite{kawabata2019symmetry}:
\begin{align}\label{10}
	\mathcal{C} H^T \mathcal{C}^{-1}=-H.
\end{align}
Therefore, Eq.~\ref{10} is used to describe the particle-hole symmetry in non-Hermitian systems.

Similarly, in non-Hermitian systems, chiral symmetry includes~\cite{kawabata2019symmetry}:
\begin{align}\label{11}
	\Gamma H \Gamma^{-1}=-H,
\end{align}
\begin{align}\label{12}
	\Gamma H^{\dagger} \Gamma^{-1}=-H,
\end{align}
where $\Gamma$ is a unitary matrix. In addition, time-reversal symmetry is an independent symmetry class in one-dimensional Hermitian systems, satisfying~\cite{ashida2020non}:
\begin{align}\label{13}
	\mathcal{T} H^* \mathcal{T}^{-1}=H,
\end{align}
where $\mathcal{T}=U_{\mathcal{T}}\mathcal{K^*}$ is a unitary matrix, and $\mathcal{K^*}$ is a complex conjugate operation. Furthermore, the time-reversal symmetry in non-Hermitian systems is expressed as~\cite{ashida2020non}:
\begin{align}\label{14}
	\mathcal{T} H^T \mathcal{T}^{-1}=H.
\end{align}

\begin{table}[htbp]
	\caption{In the presence of symmetry, the properties of eigenvalues \(E_i\) for \(2\times 2\) and \(3\times 3\) non-Hermitian matrices (Table adapted from Ref.~\cite{delplace2021symmetry}). PT denotes parity-time symmetry and CP denotes charge-conjugate parity symmetry. Degree \(n\) denotes the dimension of the system Hamiltonian, and sgn \(\Delta\) indicates symmetry (\(\Delta=+1\)) and symmetry breaking (\(\Delta=-1\)).} 
	\centering
	% 1. 设置列间距为 6pt（调整这个值可以控制表格左右铺满的程度）
	\setlength{\tabcolsep}{20pt}
	% 2. 稍微放大表格内的字体（例如 \large 或 \normalsize），使其撑满全宽 
	% 注意：这里把 tabularx 改为了普通的 tabular
	% 列格式：c 表示居中，l 表示左对齐。根据截图，后两列明显是居中或左对齐的。
	\begin{tabular}{c c l l}
	    \hline\hline  % booktabs 宏包的粗顶线
	    Degree \(n\) & sgn \(\Delta\) & PT & CP \\
		\midrule  % booktabs 的中线
		\multirow{3}{*}{2} 
		& \(+1\) & \(E_i \in \mathbb{R},\ E_1 \neq E_2\) & \(E_1 = -E_2^{*}\) \\
		& \(0\)  & \(E_1 = E_2\) & \(E_1 = E_2\) \\
		& \(-1\) & \(E_1 = E_2^{*}\) & \(E_i \in i\mathbb{R},\ E_1 \neq E_2\) \\
		\midrule  % booktabs 的分隔线
		\multirow{3}{*}{3} 
		& \(+1\) & \(E_i \in \mathbb{R},\ E_1 \neq E_2 \neq E_3\) & \(E_1 = -E_2^{*}\) and \(E_3 \in i\mathbb{R}\) \\
		& \(0\)  & \(E_1 = E_2\) and \(E_3 \in \mathbb{R}\) & \(E_1 = E_2\) and \(E_3 \in i\mathbb{R}\) \\
		& \(-1\) & \(E_1 = E_2^{*}\) and \(E_3 \in \mathbb{R}\) & \(E_i \in i\mathbb{R},\ E_1 \neq E_2 \neq E_3\) \\
		\bottomrule % booktabs 宏包的粗底线
	\end{tabular}
	\label{tab1}
\end{table}

Beyond the conventional internal symmetries previously discussed, certain non-Hermitian systems also exhibit parity-time symmetry. Initially introduced in theoretical studies, parity-time symmetric non-Hermitian systems allow for entirely real spectra. For a Hamiltonian $H(r,p,t)$, where $r$, $p$, and $t$ denote the position, momentum, and time, respectively. If $H$ is a parity-time symmetric Hamiltonian, then it satisfies the commutation relation $[H,\rm{PT}]=0$ (PT is the combined parity and time reversal operator). Here, the parity operator P performs a spatial reflection, transforming $r\rightarrow -r$ and $p\rightarrow -p$. The time reversal operator T reverses the direction of time, mapping $t\rightarrow -t$ and simultaneously flipping the sign of momentum, $p\rightarrow -p$. In particular, the emergence of a phase transition point breaks the symmetry, leading to the spontaneous breaking of parity-time symmetry beyond this point. Such a phase transition point is typically an exceptional point, where the system's eigenenergies and eigenstates coalesce simultaneously. After traversing the exceptional point, the system generates symmetric and symmetry-broken phases. In the parity-time symmetric phase, the eigenvalues have distinct real parts but degenerate imaginary parts. Conversely, in the parity-time broken phase, the real parts become degenerate while the imaginary parts split. Recently, charge-conjugate parity symmetry has also been identified in non-Hermitian systems~\cite{zhang2025exceptional,delplace2021symmetry}. A non-Hermitian Hamiltonian with parity-time symmetry protection maintains a purely real energy spectrum, whereas the behavior of energy spectra under charge-conjugate parity symmetry is more complex. Researchers have characterized the energy spectra of systems with parity-time and charge-conjugate parity symmetry or their breaking, as shown in Table \ref{tab1}.

\begin{figure}
	\centering
	\includegraphics[width=0.98\textwidth]{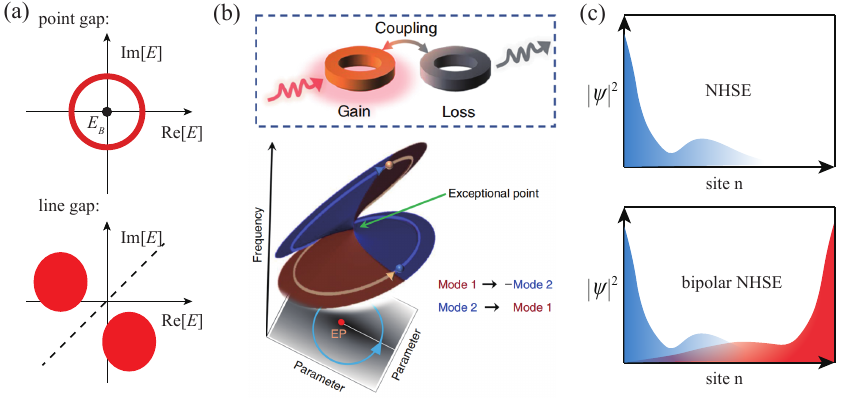}
	\caption{(a) On the complex spectral plane,  non-trivially encircle nonzero areas and form the so-called point gaps, or may eigenbands (red) may be separated by line gaps, thereby leading to much richer symmetry-protected topological states. (b) a universal photonic non-Hermitian system consists of coupled gain and loss materials, such as a ring resonator (inset in Figure A). Under certain critical coupling values and gain/loss coefficients, exceptional points may occur in the frequency spectrum of such systems.The topological properties of a second-order exceptional point are best understood by two Riemann surfaces (shown in blue and brown colours) connected at the square root branch point. Starting on the upper surface, one ends up on the lower surface after one round and vice versa (the states move smoothly from one surface to the other). Encircling twice brings one back to the initial point, but the modes acquire a $\pi$-Berry phase; after 4 cycles, the modes return to their starting phases (Figure adapted from Ref.~\cite{midya2018non}). (c) All eigenstates are localized at the system boundary in an exponentially decaying manner, which is the non-Hermitian skin effect.}   
	\label{FigR1}
\end{figure}
\subsection{ Topological properties of the non-Hermitian systems}
The topological phenomena emerging in non-Hermitian systems are indeed more diverse and abundant compared to traditional Hermitian systems. The core difference lies in the complex eigenvalues of the non-Hermitian system, which directly introduces new dimensions and symmetry requirements, thereby giving rise to a large number of abundant phenomena. Therefore, topological phases have been a major focus of research, particularly regarding their distinctive characteristics. Generally, the complex energy plane used to describe topological properties includes two types of complex energy gaps~\cite{kawabata2019symmetry}: point gap [Fig.~\ref{FigR1}(a) Upper plane] and line gap [Fig.~\ref{FigR1}(a) Lower plane]. In the presence of a point (line) gap, complex energy bands do not cross a reference point (line) in the complex energy plane. Physically, point gaps are relevant to localization transitions in one-dimensional non-Hermitian systems, whereas topologically protected edge states are often explained via line gaps in the real part of the spectrum~\cite{zhang2019correspondence,okuma2020topological}.

A non-Hermitian Hamiltonian $H$ is defined to have a point gap when its complex plane spectrum avoids a reference point $E_B\in\mathbb{C}$, i.e., det$(H-E_B)\neq0$. The simplest nontrivial example of the point gap topological phases appears in one-dimensional systems. Whereas det$(H-E_B)$ is always real for Hermitian $H$, it can be complex for non-Hermitian $H$. This leads to the following definition of the winding number $W(E)$~\cite{gong2018topological}:
\begin{align}
	W(E)=\int_0^{2\pi} \frac{d k}{2\pi\mathrm{i}}\frac{d}{d k}\log\operatorname{det}(H(k)-E_B)
\end{align}
where $H(k)$ is the non-Hermitian Bloch Hamiltonian in momentum space with a finite number of bands ($k\in[0,2\pi]$). The topological properties of point gap spectra are usually characterized by non-zero winding numbers $W(E)\neq 0$.

Non-Hermitian systems also exhibit a range of fascinating phenomena, such as exceptional points in the energy spectrum and the non-Hermitian skin effect. Exceptional points represent a special type of degeneracy where not only do the eigenvalues (both real and imaginary parts) coalesce, but the corresponding eigenvectors also become fully parallel. Near exceptional points, the eigenvalues exhibit extremely high sensitivity to small variations in system parameters. In the simplest two mode system, a perturbation parameter $\varepsilon$ to the Hamiltonian can induce eigenfrequency shifts on the order of $\sqrt{\varepsilon}$. This $\sqrt{\varepsilon}$-scaling suggests a greatly enhanced response for small $\varepsilon$, paving the way for applications in ultrasensitive sensing~\cite{sarkar2025exponentially,budich2020non,koch2022quantum,chen2017exceptional}. Moreover, in the steady-state regime, dynamically encircling a second-order exceptional point over a full cycle results in eigenstate exchange accompanied by the acquisition of a $\pi$ Berry phase~\cite{midya2018non}. This behavior originates directly from the topological structure associated with exceptional points in the complex eigenvalue spectrum (Fig.~\ref{FigR1}(b)). Another remarkable phenomenon in non-Hermitian and open quantum systems is the non-Hermitian skin effect, which describes the localization of eigenstates at boundaries~\cite{YaoShuYu2018,lee2019anatomy,xiao2020non,xiao2024restoration}, as illustrated in Fig.~\ref{FigR1}(c). The non-Hermitian skin effect plays a vital role in understanding topological phases in non-Hermitian systems~\cite{lee2018hybrid,zhang2022review}. Recent studies have extended the concept to more complex manifestations such as the bipolar non-Hermitian skin effect~\cite{SongFei2019,jiang2023reciprocating,zhao2024topological} and the many-body critical non-Hermitian skin effect~\cite{qin2025many,hu2025many}, which hold promise for applications in optical devices and quantum information technologies. 

\subsection{Lindblad master equation}
Under the Markov approximation, open systems quantum dynamics are described by the Lindblad master equation~\cite{yu2021generalized}:
\begin{align}
	\label{Eq1}
	\dot{\rho}=&-i[H, \rho]+\sum_j\left(L_j \rho L_j^{\dagger}-\frac{1}{2}\left\{L_j^{\dagger} L_j, \rho\right\}\right),
\end{align}
where $\rho$ denotes the system's density matrix, $H$ represents the Hermitian Hamiltonian, and $L_j$ is the jump operator. This can be written compactly as $d\rho(t)/dt=\mathcal{L}\rho(t)$, with $\mathcal{L}$ denoting the Liouvillian superoperator, a non-Hermitian operator that effectively governs the non-unitary evolution of the density matrix. By introducing the effective non-Hermitian Hamiltonian $H_{\mathrm{NH}} = H - \frac{i}{2} \sum_j L_j^{\dagger} L_j$, the master equation takes the form~\cite{yu2021generalized}:
\begin{align}
	\frac{\partial}{\partial t} \rho = -i\left(H_{\rm NH}\rho-\rho H_{\rm NH}^{\dagger}\right) +\sum_j L_j \rho L_j^{\dagger}.
\end{align}
Here, the first two terms capture the non-unitary evolution generated by 
$H_{\rm NH}$, while the last term represents quantum jumps due to environmental coupling. On time scales shorter than the typical interval between quantum jumps, the dynamics are approximately governed by $H_{\rm NH}$ alone. In this regime, a pure state evolves as $\partial_t |\psi\rangle = -i H_{\text{NH}} |\psi\rangle$. As time evolves, there is a finite probability of a quantum jump occurring, producing a new state $|\psi(t)\rangle \rightarrow L_j |\psi(t)\rangle$.
Subsequently, this state continues to evolve under $H_{\text{NH}}$ until the next quantum jump occurs. This process defines a quantum trajectory of the state $\ket{\psi}$ driven jointly by the non-Hermitian effective Hamiltonian and the quantum jumps. 

Experimentally, measurement devices can respond to whether the quantum transition occurs or not, and then select a certain quantum trajectory according to the measurement result, which corresponds to the post selection. Since the density matrix can be regarded as the superposition of many pure states, the evolution of the density matrix under the Lindblad quantum master equation can be obtained by summing all possible quantum trajectories. Consequently, the study of open quantum systems centers on two key non-Hermitian operators: $H_{\text{NH}}$, describing the time evolution of the wave function under post selection, and $\mathcal{L}$, describing the time evolution of the density matrix. 

\subsection{Liouvillian exceptional point in open system}
In quantum open systems, exceptional points also appear in the Liouvillian spectrum, which is described by the evolution of the density matrix driven by the Liouvillian operator~\cite{chen2022decoherence,sun2023chiral}. Hui-Xia Gao et al. utilized the flexible control characteristics of single-photon interferometry to thoroughly investigate the phenomenon of chiral state transfer when the Liouvillian exceptional point is parametrically encircled~\cite{gao2025photonic}. The study reveals that the chirality of the dynamics only manifests within an intermediate timescale for encircling and is governed by the structure of the Liouvillian spectrum near the exceptional point. Importantly, the evolutionary trajectory generated by encircling an exceptional point exhibits similarities to the features presented in the Liouvillian spectrum diagram and the Riemann surface of the non-Hermitian Hamiltonian. Additionally, their findings reveal that over longer time scales, the system will relax to a steady state, with the chirality disappearing and becoming independent of the direction of parametric encircling. The experiment explicitly confirms the existence of chiral state transfer near the Liouvillian exceptional point, revealing its transient characteristics and laying the foundation for future exploration of the rich physical implications of Liouvillian exceptional points in open quantum systems.

This study takes an open system containing two energy levels $\{\ket{0},\ket{1}\}$ as an example, and its hybrid Lindblad master equation can be described as~\cite{gao2025photonic}:
\begin{align}
	\dot{\rho} & =-i\left(H \rho-\rho H^{\dagger}\right)+L_\phi \rho L_\phi^{\dagger}-\frac{1}{2} L_\phi^{\dagger} L_\phi \rho-\frac{1}{2} \rho L_\phi^{\dagger} L_\phi:=\mathcal{L} \rho,
\end{align}
where $\mathcal{L}$ is a Liouvillian superoperator, and the non-Hermitian Hamiltonian is given by $H=H_0-i \Gamma|1\rangle\langle 1|$,
\begin{align}
	H_0=\left(\begin{array}{cc}
		\frac{\delta_1}{2} & -\Omega_1 \\
		-\Omega_1^* & -\frac{\delta_1}{2}
	\end{array}\right).
\end{align}

\begin{figure}
	\centering
	\includegraphics[width=0.98\textwidth]{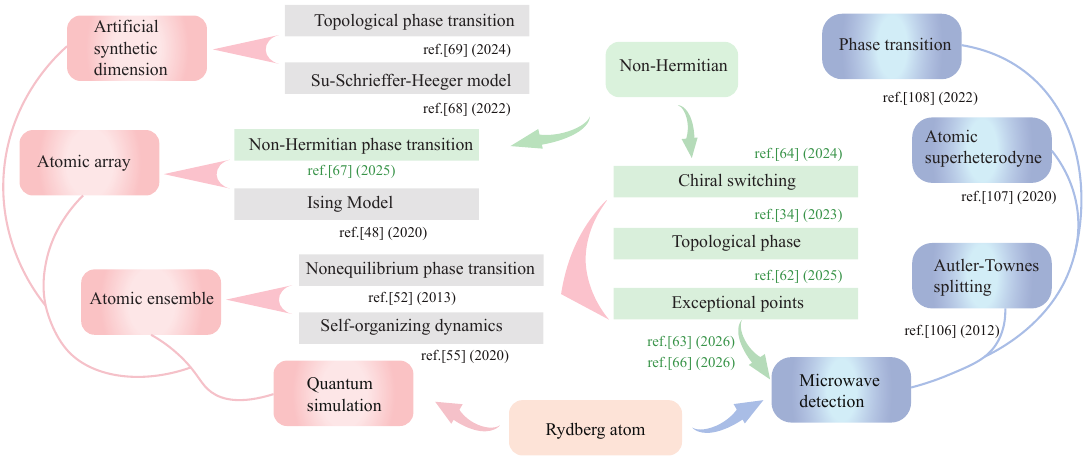}
	\caption{Development status diagram of Rydberg atomic systems.}
	\label{Fig.A2}
\end{figure}

The dephasing channel is characterized by the quantum jump operator $L_\phi=\sqrt{\gamma_{\phi}}\ket{1}\bra{1}$ with the dephasing rate $\gamma_{\phi}$. The Liouvillian eigenspectrum $\lambda_i$ is calculated through $\mathcal{L}\rho_i^{R}=\lambda_{i}\rho_i^{R}$ with the corresponding Liouvillian eigenstates $\rho_i^{R}$. Exceptional points can also arise in the Liouvillian eigenspectrum with coalescing eigenenergies and eigenstates. 
\subsection{The development status of Rydberg atomic systems}
Rydberg atoms, characterized by their strong long-range interactions, large electric dipole moments, flexible tunability of energy levels, and excellent scalability, have emerged as a powerful platform for quantum simulation and microwave sensing. In recent years, the development in this field has exhibited a clear trajectory from Hermitian to non-Hermitian frameworks. To systematically outline this evolutionary progression, we summarize the core scientific questions, developmental pathways, and application scenarios. As shown in Fig.~\ref{Fig.A2}, Rydberg atomic systems are mainly divided into atomic arrays and atomic ensembles. Atomic arrays, with their highly programmable interactions, offer unique advantages in quantum simulation~\cite{browaeys2020many}. In contrast, atomic ensembles leverage collective enhancement effects and the electromagnetically induced transparency process, excelling in quantum precision measurement~\cite{sedlacek2012microwave,jing2020atomic,ding2022enhanced} and the study of non-equilibrium physics~\cite{carr2013nonequilibrium,ding2019Phase}. Furthermore, the dense and resolvable energy level structure of Rydberg atoms provides a natural synthetic dimension~\cite{kanungo2022realizing,lu2024probing}, enabling the simulation of non-Hermitian topological models such as the Su-Schrieffer-Heeger model without the need for complex lattice fabrication, thereby opening new avenues for topological physics research.

In the field of non-Hermitian many-body physics, significant experimental progress has been achieved using Rydberg atomic systems. Jun Zhang et al. utilized cold Rydberg atomic gases to experimentally observe interaction-induced exceptional points and non-Hermitian hysteresis trajectories~\cite{zhang2025exceptional}. Ya-Jun Wang et al. introduced microwave field control to achieve quantum-enhanced microwave metrology based on flipping hysteresis trajectories, with a sensitivity surpassing the standard quantum limit~\cite{wang2026quantum}. Chong-Wu Xie et al. observed chiral switching in non-Hermitian many-body steady states in a hot atomic vapor~\cite{xie2025chiral}. Chao Liang et al. exploited the nonlinear effect near the exceptional point to achieve microwave electric field measurement with a sensitivity as high as 22.68 $\rm{nV~ cm^{-1}Hz^{-1/2}}$~\cite{liang2026exceptional}. In the realm of programmable optical tweezer arrays, Yao-Wen Zhang et al. constructed a non-Hermitian XY spin chain model and directly observed parity-time symmetry breaking as well as interaction-driven exceptional point shifts~\cite{zhang2025observation1}. Dong-Dong Hao et al. experimentally realized a dissipative version of the Su-Schrieffer-Heeger model~\cite{hao2023topological}. The diverse types of Rydberg atom platforms (whether atomic ensembles or atomic arrays), each bringing their respective strengths into full play, occupy an important position in the research of non-Hermitian quantum many-body physics.

\section{Non-Hermitian properties in Rydberg atomic systems}\label{sec2}
Rydberg atomic systems provide a highly controllable platform for studying non-Hermitian physics, where many-body interactions naturally introduce effective dissipation channels~\cite{busche2017contactless,dudin2012strongly,bariani2012dephasing}. One prominent mechanism arises from the long-range interactions between atoms excited to Rydberg states. When multiple atoms occupy Rydberg levels, exchange and collision processes lead to dissipation, which effectively generates non-Hermitian characteristics. This dissipation also originates from the long-range interactions between the Rydberg polaritons. A Rydberg polariton is a coherent superposition of a photonic component and a collective Rydberg spin wave, where the collective spin wave behaves as a Rydberg superatom. Due to the Rydberg blockade effect, once an atom is excited to a Rydberg state, it suppresses further excitations within a characteristic blockade radius around it, thereby forming a Rydberg superatom. Under electromagnetically induced transparency conditions, this Rydberg superatom can store a photon, giving rise to a Rydberg polariton~\cite{ningyuan2016observation, ripkarydberg2016}. The spatial inhomogeneity of such polaritons leads to position-dependent phase shifts, which effectively accelerate the decoherence of the Rydberg collective excitation and produce energy dissipation throughout the atomic ensemble. These interaction-induced dissipative processes enable the realization of a tunable non-Hermitian Hamiltonian in Rydberg systems, thereby facilitating investigations into exceptional points, symmetry breaking, non-equilibrium dynamics, and topological phenomena in open quantum systems~\cite{zhang2025observation,zhang2025exceptional,wang2026quantum,xie2025chiral,vsumarac2026controlling,liang2026exceptional}.

\subsection{Electromagnetically induced transparency and Rydberg polaritons}
Electromagnetically induced transparency is a quantum interference effect that makes an otherwise opaque atomic medium transparent to a resonant probe field. One implementation employs a ladder-type three-level system consisting of a ground state $|g\rangle$, an intermediate state $|e\rangle$, and a highly excited Rydberg state $|r\rangle$. A weak probe field couples the $|g\rangle \leftrightarrow |e\rangle$ transition with Rabi frequency $\rm{\Omega_p}$ and detuning $\rm{\Delta_p}$, while a strong control field drives the $|e\rangle \leftrightarrow |r\rangle$ transition with Rabi frequency $\rm{\Omega_c}$ and detuning $\rm{\Delta_c}$. Under two-photon resonance conditions ($\rm{\Delta_p} + \rm{\Delta_c}=0$), destructive quantum interference between the two excitation pathways opens a transparency window for the probe field. A Rydberg polariton is a coherent superposition of a photon and a collective Rydberg spin wave under electromagnetically induced transparency conditions. Rydberg polaritons store a photon in a collective Rydberg excitation, which can
be described as:
\begin{align}
	\ket{\rm{R_{polariton}}}=\alpha \ket{G}\ket{0} + \beta \ket{W}\ket{1},
\end{align}
where $|G\rangle = |gg\ldots gg\rangle$ is the ground state of the Rydberg superatom, $|W\rangle = \frac{1}{\sqrt{N}}\sum_i |g\ldots r_i \ldots g\rangle$ is the collective excited state (usually referred to as the Rydberg superatom), and $|0\rangle$ and $|1\rangle$ represent the Fock states of the optical field. When multiple polaritons coexist in the atomic medium, the Rydberg excitations they carry interact with each other via the van der Waals interaction ($V_{\mathrm{vdW}} \propto {C_6}/{R^6}$, where $C_6$ is the dispersion coefficient and $R=|r_i-r_j|$ is the distance between the two Rydberg atoms). The strong van der Waals interaction between Rydberg polaritons induces a spatially dependent phase shift $V_{jk}$ during the interaction time. This non-uniform phase accumulation dephases the collective Rydberg excited state, manifesting as an effective dissipation. In the non-Hermitian framework, this dissipation is captured by an imaginary on-site potential $i\gamma_{\text{eff}}$ in the effective Hamiltonian, where $\gamma_{\text{eff}} = V\rho_{rr}$ depends on both the interaction strength $V$ and the Rydberg population $\rho_{rr}$. Thus, the interaction-induced phase shift is the origin of dissipation, and both phenomena are unified under the same non-Hermitian description.

\subsection{Non-Hermitian hysteresis trajectories induced by many-body interactions}
\begin{figure}
	\centering
	\includegraphics[width=0.98\textwidth]{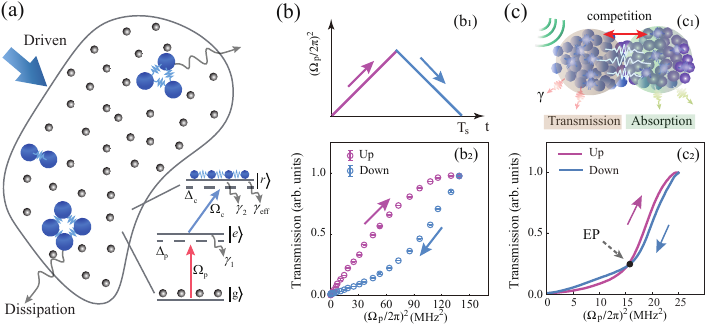}
	\caption{Schematic of many-body interaction induced hysteresis loops in Rydberg atoms. (a) Rydberg atomic energy level diagrams. Probe $\rm{\Omega_{p}}$ and coupling $\rm{\Omega_{c}}$ fields excite atoms with detunings ${\rm{\Delta_{p}}}$ and ${\rm{\Delta_{c}}}$. $\gamma_{1}$ and $\gamma_2$ are decay rates of states $\ket{e}$ and $\ket{r}$, and $\gamma_{\text{eff}}$ is the decay rate caused by Rydberg many-body interactions. (b) Measured transmission by positively (Up, pink) and negatively (Down, blue) scanning $(\rm{\Omega_{p}}/2\pi)^{2}$ in ($\rm{b}_1$), the trajectories connected by data points exhibit hysteresis loop in ($\rm{b}_2$) (Figure adapted from Ref.~\cite{zhang2025exceptional}). (c) The atomic system in ($\rm{c}_1$) and hysteresis trajectory in ($\rm{c}_2$) after adding microwaves field (Figure adapted from Ref.~\cite{wang2026quantum}). Competition arises between transmission and absorption in the electromagnetically induced transparency spectrum, and eventually the intersecting hysteresis trajectories give rise to an exceptional point (EP).}
	\label{FigA3}
\end{figure}
Based on the three-level Rydberg atomic system shown in Fig.~\ref{FigA3}(a), the system consists of three atomic manifolds: the ground state $\ket{g}$, metastable state 
$\ket{e}$, and Rydberg state $\ket{r}$. The probe field drives the transition $\ket{g} \rightarrow \ket{e}$ with Rabi frequency (detuning) $\rm{\Omega_p}$ ($\rm{\Delta_p}$), while the coupling field drives the transition $\ket{e} \rightarrow \ket{r}$ with Rabi frequency (detuning) $\rm{\Omega_c}$ ($\rm{\Delta_c}$). The spontaneous decay rates of states $\ket{e}$ and $\ket{r}$ are denoted by $\gamma_1$ and  $\gamma_2$, respectively, and $\gamma_{\rm{eff}}$ represents the effective decay rate induced by Rydberg many-body interactions. For a pair of atoms $i$ and $ j$ located at positions $r_i$ and $r_j$, when both are excited to the Rydberg state $\ket{r}$, they interact via a van der Waals (vdW) potential. Under the rotating-wave approximation, the Hamiltonian of the system can be written as~\cite{zhang2025exceptional}:
\begin{equation}
	\begin{aligned}
		H_{many} &=
		\sum_{j=1}^{N} \left[-({\rm{\Delta_p}}+{\rm{\Delta_c}})\hat{\sigma}_{rr}^{j}-\left({\rm{\Omega_p}}\sigma_{eg}^j+{\rm{\Omega_c}}\sigma_{re}^j +{\rm H.c.}\right)\right]\\
		&+\sum_{j=1}^{N} \left[-{\rm{\Delta_p}}\hat{\sigma}_{ee}^j +\sum_{j<k}V_{jk}\sigma_{rr}^j\sigma_{rr}^k\right],
	\end{aligned}
\end{equation}
where $\sigma^{j}_{\alpha\beta}=|\alpha_j\rangle\langle \beta_j|$ ($\alpha, \beta = e, g, r$). 

In the mean-field approximation, a single atom is considered to experience an effective field resulting from its interactions with all other atoms. Consequently, the problem of solving the dynamic many-body system is reduced to addressing the dynamics of a single atom within the effective field, while the other atoms are treated as part of the environment. In the following treatment, the reduced density matrix of the single atom system is used, and the non-Hermitian Hamiltonian is given by~\cite{zhang2025exceptional}:
\begin{equation}
	\begin{aligned}
		H_{1} &=-{\rm{\Delta_p}}\hat{\sigma}_{ee}
		-({\rm{\Delta_p}}+{\rm{\Delta_c}}-\frac{i \gamma_{\rm eff}}{2})\hat{\sigma}_{rr}\\
		&-\left({\rm{\Omega_p}}\sigma_{eg}+{\rm{\Omega_c}}\sigma_{re}+{\rm H.c.}\right),
	\end{aligned}
\end{equation}
where $\gamma_{\rm eff}$ reveals the effective decay rate induced by the interaction with the environment. When ${\rm{\Delta_p}}={\rm{\Delta_c}}=0$, the non-Hermitian Hamiltonian is charge-conjugate parity symmetric, and it satisfies ${U_{CP}}{H_1}U_{CP}^{-1}=-{{H_1}^*}$ with ${U_{CP}}={\rm{diag}}(1,-1,1)$ \cite{zhang2025exceptional,delplace2021symmetry}. 

Jun Zhang et al. have experimentally observed non-Hermitian hysteresis trajectories in a three level Rydberg atomic many-body system~\cite{zhang2025exceptional}, as shown in Fig.~\ref{FigA3}($\rm{b}_2$). The study found that the hysteresis trajectories formed during the upward and downward scans of the probe field do not coincide, indicating the presence of interaction-induced dissipation in the system. Additionally, their study also demonstrates that introducing microwave field control to the three-level Rydberg atomic system leads to intersecting hysteresis trajectories, as illustrated in Fig.~\ref{FigA3}($\rm{c}_2$). This occurs because the introduction of the microwave field breaks the original three level structure, and the resulting four level system satisfies particle-hole symmetry. When the probe field scans past the exceptional point, the symmetry of the system is spontaneously broken, and the spectrum transitions from transmission to absorption, generating intersecting trajectories.

\begin{figure}
	\centering
	\includegraphics[width=0.8\textwidth]{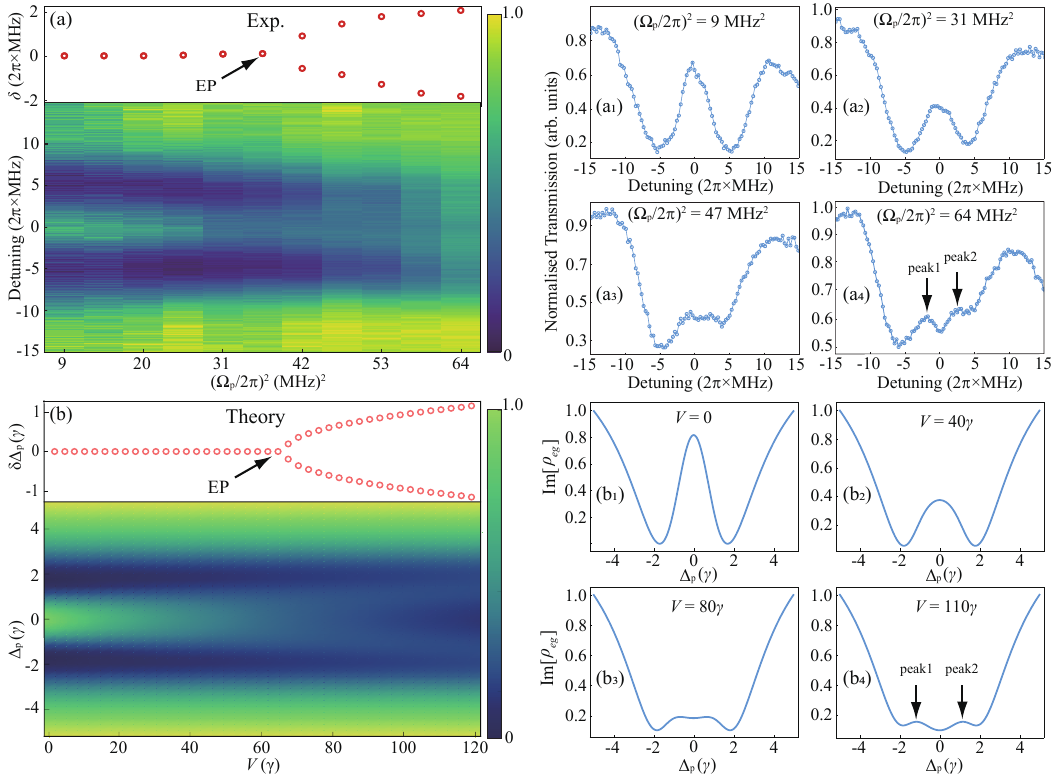}
	\caption{Measured phase diagram and theoretical phase diagram (Figure adapted from Ref.~\cite{zhang2025exceptional}). (a) The measured phase diagram versus the probe intensity $(\rm{\Omega_{p}}/2\pi)^{2}$ and detuning $\rm{\Delta_p}$. ($\rm{a_1}$)-($\rm{a_4}$) The measured electromagnetically induced transparency spectrum varies with the intensity of the probe field. (b) Theoretical spectrum Im$[\rho_{eg}]$ versus the interaction strength $V$ and detuning $\rm{\Delta_p}$. With further increase in interaction strength, the system transits an exceptional point (marked by the black arrow in the up panel of (b)). As the interaction strength $V$ gradually increases, the spectrum Im$[\rho_{eg}]$ as shown in ($\rm{b_1}$)-($\rm{b_4}$). }
	\label{FigA4}
\end{figure}
\subsection{Exceptional points in cold Rydberg atomic gases}
Recent studies have shown that open systems undergo phase transitions at exceptional points, leading to a variety of interesting physical phenomena, including chirality~\cite{shu2024chiral,wang2020electromagnetically}, unidirectional transmission or reflection~\cite{huang2017unidirectional,du2022light}, topological phase transitions~\cite{ding2024electrically,gong2024topological}, symmetry breaking~\cite{el2018non,delplace2021symmetry,li2023exceptional}, and high sensitivity detection~\cite{rosa2021exceptional,yang2023spectral}. These properties have become the focus of research on non-Hermitian systems related to exceptional points, driving a series of experimental studies in optics, electronics, and cold atomic physics. The combination of non-Hermitian properties and many-body interactions makes the emergence of non-trivial effects possible, thus providing a platform for the study of emerging phases.

Jun Zhang et al. experimentally demonstrated the emergence of exceptional points and associated spectral bifurcations in cold Rydberg atomic gases~\cite{zhang2025exceptional}. By measuring the transmission spectrum as a function of probe detuning $\rm{\Delta_{p}}$ and intensity $(\rm{\Omega_{p}}/2\pi)^{2}$, a phase diagram was mapped, revealing the process of crossing the exceptional point as the probe field intensity increases [Fig. \ref{FigA4}(a)]. Under weak probe intensity, the system exhibits a conventional electromagnetically induced transparency spectrum [Fig. \ref{FigA4}(a$_1$)]. As the probe intensity increases, interactions between Rydberg atoms are enhanced, resulting in stronger dissipation. Consequently, the electromagnetically induced transparency peak weakens [Fig. \ref{FigA4}(a$_3$)], and eventually splits into two distinct peaks at higher intensities [Fig. \ref{FigA4}(a$_4$)], signifying the crossing of a second-order exceptional point. The peaks in Fig.~\ref{FigA4}(a$_4$) are asymmetric due to the small shift in the Rydberg energy level. Moreover, the observed splitting beyond the exceptional point serves as a signature of the breaking of charge-conjugate parity symmetry~\cite{delplace2021symmetry}. 

Theoretically, by solving the master equation Eq.~\ref{Eq1} under steady state conditions $\dot{\rho}=0$, they simulated the spectrum of Im$[\rho_{eg}]$ as a function of interaction strength $V$ and the detuning ${\rm{\Delta_p}}$ [Fig. \ref{FigA4}(b), lower panel]. Similar to the experimental observations, as the interaction $V$ increases from $V=0$ to $V=110\gamma$, the peak of Im$[\rho_{eg}]$ decreases, and two peaks emerge, as illustrated in Figs. \ref{FigA4}($\rm{b}_1$)-($\rm{b}_4$) and the upper panel of Fig. \ref{FigA4}(b). During this process, the system passes through exceptional points, and the peak of Im$[\rho_{eg}]$ splits into two, indicating that a degenerate complex eigenvalue of the system has become non-degenerate.

\subsection{Chiral switching of many-body steady states in a dissipative Rydberg gas}
Non-Hermitian physics provides a unique perspective and insights into exploring the non-equilibrium dynamics of dissipative Rydberg gases. Chong-Wu Xie et al. experimentally observed the chiral switching of many-body steady states in dissipative hot Rydberg vapors~\cite{xie2025chiral}. These phenomena are similar to the typical chiral state transfer observed in non-Hermitian systems~\cite{chen2022decoherence,sun2023chiral,chen2021quantum}. This research reveals that the existence of both bistability and chiral dynamics originate from the Liouvillian exceptional structure, where two exceptional lines merge at a higher-order exceptional point. As parameters undergo adiabatic variation along a closed path, the Rydberg gas can switch between two distinct collective steady states characterized by different Rydberg excitations and optical transmissions. From the perspective of non-Hermitian physics, dissipative Rydberg gases establish a paradigm for chiral mode switching via parameter tuning and open up application avenues based on unique non-Hermitian properties.

\begin{figure}
	\centering
	\includegraphics[width=0.8\textwidth]{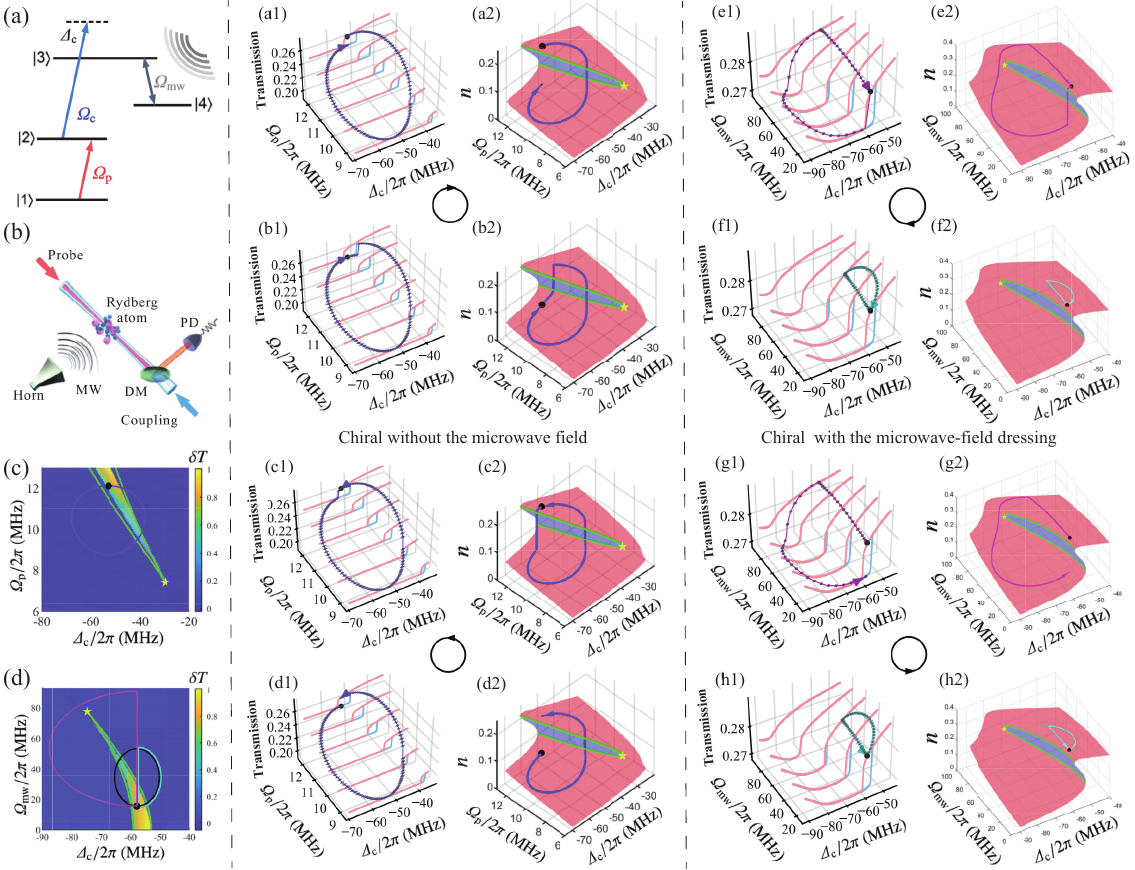}
	\caption{Exceptional structures in a thermal Rydberg gas and the measured chiral mode (Figure adapted from Ref.~\cite{xie2025chiral}). (a) Schematics of the coupling scheme. (b) Experimental setup. (c)-(d) Experimentally measured phase diagrams. The color bar represents the normalized transmission difference, $\delta T$ between the forward and backward scans of ${\rm{\Delta_c}}$. The green lines represent the numerically fitted second-order exceptional lines, and the yellow star is the third order exceptional point. (a1-d1) Chiral mode switching without the microwave field. (a2-d2) The corresponding numerically calculated encircling trajectories on the landscape of the steady-state Rydberg population. (e1-h1) Chiral mode switching with the microwave-field dressing. (e2-h2) The corresponding numerically calculated encircling trajectories on the landscape of the steady-state Rydberg population.}
	\label{FigA5}
\end{figure}
Similar to the formation of exceptional point rings in non-Hermitian physics~\cite{zhang2018dynamically}, the choice of the initial state is crucial for observing chiral mode switching. In this experiment, the authors initialize the system in a steady state within the bistable region and slowly vary it along the purple loop shown in Fig.~\ref{FigA5}(c) without the microwave field dressing and in Fig.~\ref{FigA5}(d) with the microwave field dressing. Figures~\ref{FigA5}(a1) and (c1) show the measured transmittance when the system is initialized in a high-energy steady state without microwave dressing. When the parameters are varied clockwise, the measured transmittance decreases upon returning to the initial point, indicating a state switch. Conversely, when the parameters are rotated counterclockwise, the measured transmittance returns to its initial value, and the system reverts to the original steady state. As illustrated in Figs.~\ref{FigA5}(b1) and (d1), when the system is initialized in a low-energy steady state, the chirality of the state switching is reversed. In the case of microwave dressing, richer chiral switching phenomena are demonstrated in Figs.~\ref{FigA5}(e1)-(h1), with all theoretical simulation results presented in Figs.~\ref{FigA5}(a2)-(h2).

\subsection{Controlling Rydberg atom-polariton interactions: from exceptional points to fast
	readout}
The long-range interactions and tunability of Rydberg atoms have become the foundation for many frontier fields, such as quantum computing, simulation, and sensing. When coupled with the photons propagating in the atomic cloud, they can achieve optical nonlinear effects, forming collective atomic excitation states known as Rydberg polaritons. These polaritons have strong interactions with each other. Tamara $\rm{\check{S}}$umarac et al. experimentally investigated the interactions between Rydberg polaritons in an atomic ensemble and a neighboring individual Rydberg atom~\cite{vsumarac2026controlling}, revealing three distinct quantum dynamical regimes of transmission: polariton blockade, coherent exchange, and probabilistic hopping. The exceptional point acts as the critical transition between the polariton blockade and coherent exchange mechanisms. This research reveals the physical mechanism of the dipole interaction between Rydberg atoms and polaritons, laying the foundation for readout techniques based on ensembles in neutral atomic quantum processors and nonlinear photon networks for photon quantum random walks~\cite{schreiber20122d}.

\begin{figure}
	\centering
	\includegraphics[width=1\textwidth]{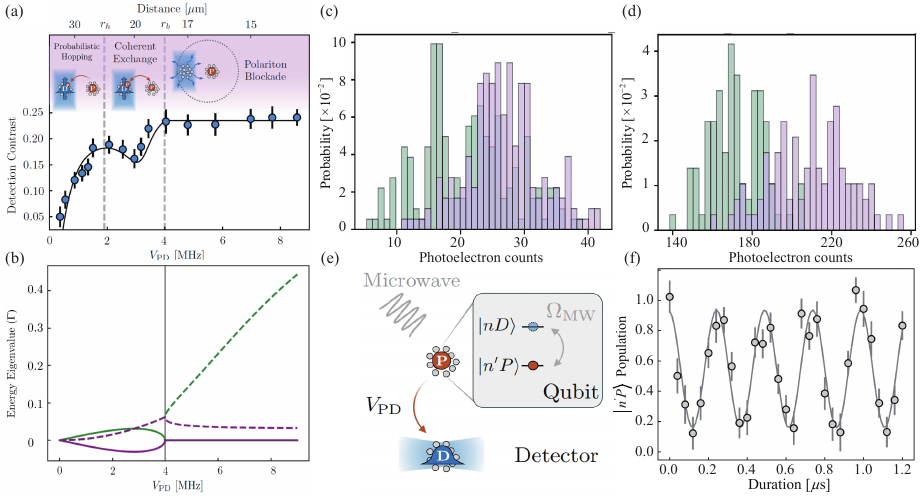}
	\caption{Rydberg electromagnetically induced transparency as a remote detector of single Rydberg atoms (Figure adapted from Ref.~\cite{vsumarac2026controlling}). (a) Three distinct regimes, corresponding to probabilistic hopping, coherent exchange, and polariton blockade, can be identified. (b) The real (solid) and imaginary (dashed) parts of the eigenvalues as a function of the dipole-dipole interaction strength, $V_\mathrm{PD}$. An exceptional point clearly emerges at the interaction strength corresponding to the blockade radius (gray line). (c) Single-shot histogram for 30 $\mu s$ of readout, showing the difference in detected photons exiting the detector ensemble with (green) and without (purple) a Rydberg atom nearby. The measurement fidelity is 47(5) \%. (d) Using repeated preparation and detection (20 repetitions of 10 $\mu s$ detection), we achieve a fidelity of 79(4) \%. (e) Detection schematic diagram. (f) Following a global microwave drive, state-selective and remote detection of the Rydberg qubit is performed, showing long-lived coherent microwave Rabi oscillations.}
	\label{FigA6}
\end{figure}

As shown in Fig.~\ref{FigA6}(a), the contrast exhibits clear piecewise characteristics with the variation of the atom-polariton distance. Meanwhile, the eigenvalue calculations of a non-Hermitian Hamiltonian in Fig.~\ref{FigA6}(b) reveal an exceptional point emerging between the blockade and coherent exchange regimes. Furthermore, the authors utilized Rydberg atom-polariton interactions to conduct experiments on fast, remote, and non-destructive qubit readout. Figure \ref{FigA6}(c) presents the results of a single-shot detection of a Rydberg atom, achieving a fidelity of 47(5)\%. Figure \ref{FigA6}(d) also shows that repeated detection improved the fidelity to 79(4) \%. Additionally, Figures \ref{FigA6}(e) and (f) demonstrate the capability of this remote detection method for state-selective and remote detection of a single Rydberg qubit over a distance of 20 $\mu \rm{m}$. After applying a global microwave field drive, long-lived and coherent microwave Rabi oscillations were observed, confirming the practical applicability of this approach.

\subsection{Collective dissipation engineering of interacting Rydberg atoms}
In recent years, researchers have been able to precisely control the interactions between atoms and the loss conditions at the microscopic scale within the Rydberg atomic array. Tao Chen et al. combines the single-particle control capabilities of Rydberg atom arrays with laser-induced tunable dissipation channels, providing a powerful platform for exploring strongly correlated open quantum systems~\cite{chen2025collective}. Their study not only reveals how interactions shift exceptional points in non-Hermitian systems, but also demonstrates interaction-enabled selective protection of spin states via the quantum Zeno effect. These results lay the foundation for future investigations of non-Hermitian many-body physics, dissipative phase transitions, and error correction in larger-scale Rydberg atom arrays.
\begin{figure}
	\centering
	\includegraphics[width=1\textwidth]{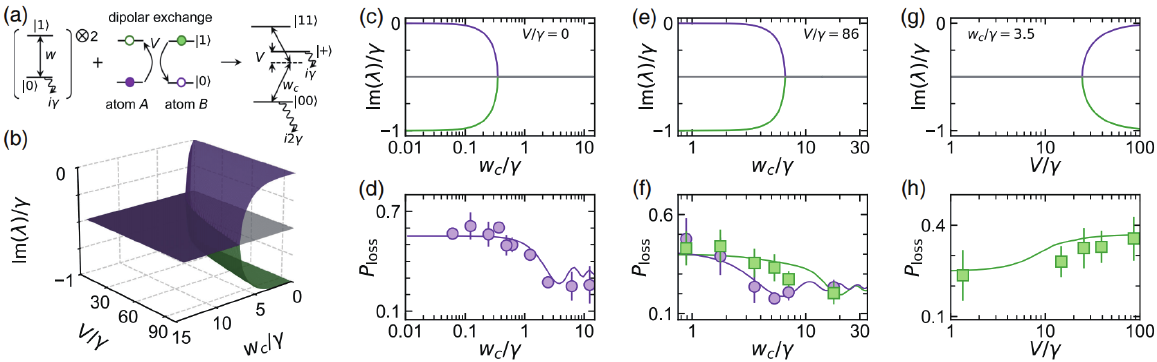}
	\caption{Experimental realization of the interacting two-body dissipative spin model (Figure adapted from Ref.~\cite{chen2025collective}). (a) This system can be cast as an effective three-level model within a pair basis, where the inherent decay rates $\Gamma_R$ of the Rydberg states $\ket{0}$ and $\ket{1}$ are ignored as $\Gamma_R \ll \gamma$ in the experiments. (b) Imaginary parts of the eigenenergies $\lambda$ of the effective two-body non-Hermitian Hamiltonian under different $V/\gamma$ and $\rm{w_c}/\gamma$ ratios. (c), (e) ${\rm{Im}}(\lambda)$ vs the $\rm{w_c}/\gamma$ ratio for (c) noninteracting $V=0$ and (e) $V/\gamma=86$. (g) ${\rm{Im}}(\lambda)$ vs the $V/\gamma$ ratio for $\rm{w_c}/\gamma=3.5$. (d), (f), (h) Normalized measured population loss after $1$~$\mu s$ time evolution for the $V/\gamma$ and $\rm{w_c}/\gamma$ ratios shown in (c),(e),(g) respectively. In (f), data for the noninteracting case are shown (purple circles) for the same parameters. Solid lines in (d), (f), (h) indicate the simulation results from the Lindblad equation with state preparation and measurement infidelity included. 
		Error bars are standard errors from multiple independently measured datasets.}
	\label{FigA7}
\end{figure}

The research team constructed a controlled dissipative two-level system. As shown in Fig.~\ref{FigA7}(a), they encoded two Rydberg states of the atoms as $\ket{0}=\ket{42 S_{1 / 2}, m_J=1 / 2}$ and $\ket{1}=\ket{42 P_{3 / 2}, m_J=1/2}$. The $\ket{42 P_{3 / 2}}$ state has a laser-induced effective loss. The states $\ket{0}$ and $\ket{1}$ were resonantly coupled by a microwave field with coupling strength $\rm{w}$. For an atom pair separated by a distance $R_{AB}$, there also exists a dipolar exchange interaction $V=C_3 /\left(2 R_{A B}^3\right)$, where $C_3$ is the dipolar interaction coefficient. Under the basis $\{\ket{00}, \ket{+}, \ket{11}\}$
where $|+\rangle=(\ket{01}+\ket{10})/\sqrt{2}$ (with antisymmetric state
$\ket{-}=(\ket{01}-\ket{10})/\sqrt{2}$ fully decoupled), the effective non-Hermitian Hamiltonian can be written as:
\begin{equation}
	\begin{aligned}
		H=\left(\begin{array}{ccc}
			-i \gamma & \rm{w_c} & 0 \\
			\rm{w_c}  & V-i \gamma / 2 & \rm{w_c} \\
			0 & \rm{w_c}  & 0
		\end{array}\right),
	\end{aligned}
\end{equation}
with the collective microwave coupling rate $\rm{w_c}=\sqrt{2}w$. Fig.~\ref{FigA7}(b) illustrates the imaginary part of the eigenvalue ${\rm{Im}}(\lambda)$ under different microwave coupling strengths and interaction strengths. The core feature of this non-Hermitian Hamiltonian is the existence of exceptional points, which partition the parameter space into parity-time symmetric and broken phases. In the symmetric phase, the imaginary parts (i.e., decay rates) of all eigenmodes are the same, and the system exhibits oscillatory behavior; in the broken phase, the decay rates of different modes split.

The experimental results are shown in Figs.~\ref{FigA7}(d), (f), and (h). For the non-interacting atom pair ($V=0$), as shown in Fig.~\ref{FigA7}(d), when the microwave coupling strength $\rm{w_c}$ is small, the system resides in the broken phase and exhibits large loss. As $\rm{w_c}$ increases and crosses the exceptional point (located at $\rm{w_c}/\gamma=\sqrt{2}/4$), the system enters the parity-time symmetric phase, where the loss is significantly suppressed. At this point, the quantum Zeno effect emerges and hinders the dissipation process. However, introducing a strong interaction $V=86\gamma$ significantly moves the exceptional point to $\rm{w_c}/\gamma \propto 7$ position [see Fig.~\ref{FigA7}(e)]. By comparing the loss behaviors of single atoms and interacting atomic pairs, it is demonstrated that the interaction can enhance the dissipation before passing through the exceptional point. In Fig.~\ref{FigA7}(h), when the interaction strength is increased in the experiment, the loss increases, which provides direct evidence for the enhanced two-body dissipation.

\section{Sensing based on exceptional points in the Rydberg atomic system}\label{sec3}
Rydberg atoms excited to high energy levels possess characteristics such as strong polarizability and significant electric dipole moments, making them highly sensitive probes for detecting electric fields~\cite{liu2022deep,zhang2024early,zhang2024ultra}. Currently, microwave detection technology based on Rydberg atoms has become a frontier in the field of quantum precision measurement~\cite{schlossberger2024rydberg,zhang2024rydberg}. Early experiments achieved high-precision measurement of microwave electric fields by combining electromagnetically induced transparency with the Autler-Townes effect~\cite{sedlacek2012microwave,sedlacek2013atom}. Dong-Sheng Ding et al. also utilized the Rydberg atomic system to achieve high sensitivity measurements from single-body to many-body phase transitions\cite{ding2022enhanced}. More recently, dissipative Rydberg atom systems exhibiting non-Hermitian properties have been shown to offer significant advantages for microwave detection~\cite{wang2026quantum,liang2026exceptional}. In particular, sensing enhanced by exceptional points in non-Hermitian systems has been demonstrated across various platforms, including optical cavities~\cite{chen2017exceptional,hodaei2017enhanced}, photonic crystals~\cite{park2020symmetry}, circuit systems~\cite{kononchuk2022exceptional,li2023stochastic}, and atomic systems~\cite{liang2023observation}. By leveraging a non-Hermitian Rydberg atomic platform, Ya-Jun Wang et al. achieved a significant enhancement in sensitivity by utilizing the nonlinear effects generated through microwave manipulation of exceptional points~\cite{wang2026quantum}. In a parallel approach, Chao Liang et al. focused on a different mechanism: they leveraged the square-root amplification of perturbations near the exceptional point, also controlled by microwaves, to enhance measurement sensitivity~\cite{liang2026exceptional}.

\subsection{Quantum enhanced metrology based on flipping trajectory of cold Rydberg gases}
\begin{figure}
	\centering
	\includegraphics[width=1\textwidth]{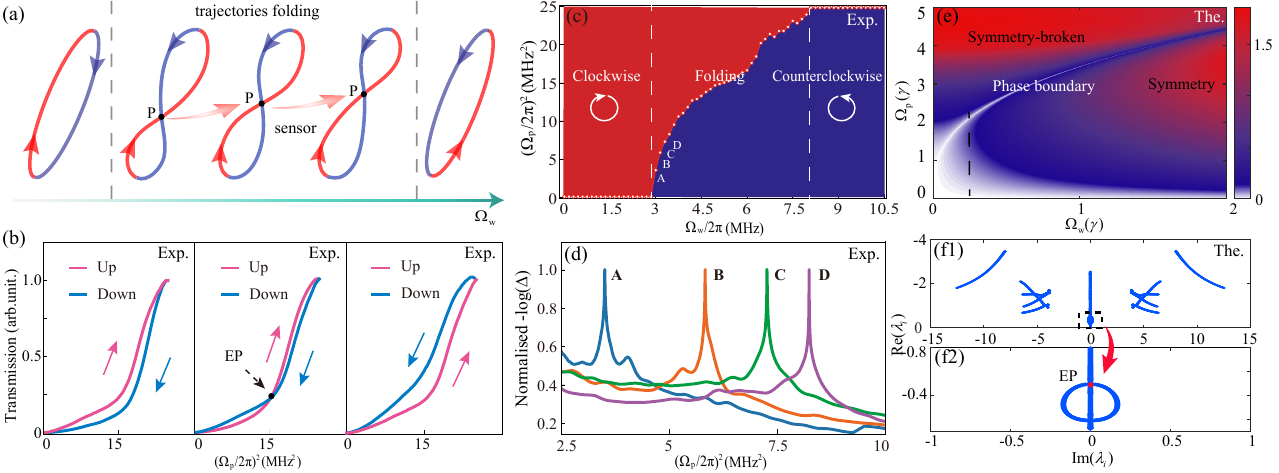}
	\caption{Measured phase diagram of hysteresis trajectories and energy gap phase diagram (Figure adapted from Ref.~\cite{wang2026quantum}). (a) Physical schematic diagram. (b) Hysteresis trajectories with microwave field reversal. (c) The measured phase diagram of hysteresis trajectories. The red and blue regions indicate $\rm{T_{up}>T_{down}}$ and $\rm{T_{up} \leq T_{down}}$, respectively. (d) The measured trajectories difference $-\rm{log(\Delta T})$, where blue, orange, green and purple correspond to the data at point A, B, C, and D in (c), respectively. (e) The energy gap is obtained by numerically solving the modulus of $E_2-E_3$, where the legend represents the magnitude of the energy gap. The white curve with a zero energy gap separates the particle-hole symmetric phase from the symmetry-broken phase. ($\rm{f}_1$)-($\rm{f}_2$) The Liouvillian eigenspectrum complex planes of the system in ($\rm{f}_1$) and the enlarged view in ($\rm{f}_2$), where an Liouvillian exceptional point emerges in the complex planes. }
	\label{FigA8}
\end{figure}

Rydberg atomic systems have exhibited rich hysteretic trajectories that arise from interaction-induced non-Hermitian properties~\cite{carr2013nonequilibrium}. Building on the previous discovery of interaction-induced exceptional points and hysteresis trajectories in cold Rydberg gases~\cite{zhang2025exceptional}, Ya-Jun Wang et al. further investigates quantum sensing based on exceptional points~\cite{wang2026quantum}. The interplay between the unique non-Hermitian dynamics of Rydberg atom systems and their complex critical properties offers promising opportunities for microwave field sensing applications~\cite{wang2026quantum}. When a microwave field is applied to the interacting Rydberg atoms, it induces a change in the energy spectrum, leading to the closure of energy gaps between different eigenstates. This `gap-closing' effect results in the emergence of `folded' hysteresis trajectories and the transition between distinct phases. In this scenario, the non-Hermitian properties undergo a dramatic change in response to the microwave field, thereby serving as a valuable resource for demonstrating enhanced sensing. Under the external microwave field driving, the hysteresis trajectories exhibit distinct folding patterns as the Rabi frequency of the probe field is increased and decreased, leading to the formation of microwave field dependent intersection points [Fig.~\ref{FigA8}(a)]. By analyzing the shifts in the intersection position in response to the external microwave field, the authors achieved high precision measurement of the external field strength, with a sensitivity surpassing the standard quantum limit. 

Experimentally, the system's dynamic evolution is monitored by measuring probe transmission. The probe intensity is modulated using a triangular waveform generated by an acousto-optic modulator with a scanning period of \(\rm{T_s}\), and the transmission trajectories during both the Up and Down processes are recorded [Fig.~\ref{FigA8}(b)]. Due to the interactions between Rydberg atoms, the interaction-induced dissipation plays a vital role in the system's dynamic evolution. The increase and decrease of Rydberg atoms [by scanning $(\rm{\Omega_{p}}/2\pi)^2$] result in asymmetrical responses to probe transmission, generating a hysteresis loop \cite{zhang2025exceptional}. The measured hysteresis trajectories map the full dynamics onto the parameter space of \(\rm{(\Omega_{p}/2 \pi)^2}\) and \(\rm{\Omega_{w}/2 \pi}\), yielding a phase diagram where regions of $\rm{T_{up}>T_{\rm{down}}}$ (red) and $\rm{T_{up} \leq T_{\rm{down}}}$ (blue) are separated by boundaries defined by the white intersection points in Fig.~\ref{FigA8}(c). The differences in hysteresis trajectories $\Delta \rm{T}= |\rm{T_{up}-T_{down}}|$ at several positions (A, B, C, D) are shown in Fig.\ref{FigA8}(d). As the transmission difference spectra have high discrimination, the intersection points of the trajectories under different microwave fields can be completely separated. In the region of hysteresis trajectory folding, the nonlinear response of the intersection points to the microwave field $\rm{\Omega}_{\rm{w}}$ offers an approach to microwave field measurement. 

Theoretically, a concise description of open quantum systems is provided by the Lindblad quantum master equation~\cite{wang2026quantum}:
\begin{align}
	\dot{\rho}=&-i[H, \rho]+\sum_j\left(L_j \rho L_j^{\dagger}-\frac{1}{2}\left\{L_j^{\dagger} L_j, \rho\right\}\right):=\mathcal{L}\rho, 
\end{align}
\begin{align}
	H=&-\sum\limits_j [2{\Delta _1}\sigma _{ee}^j + (2\Delta_{r_1}+iV_{\rm{eff}})\sigma _{r_{1}r_{1}}^j + {2\Delta _{r_2}}\sigma _{r_{2}r_{2}}^j] \nonumber\\ 
	&+\sum\limits_j [({{\rm{\Omega} _p}\sigma _{ge}^j + {\rm{\Omega _c}}\sigma _{er_1}^j + {\rm{\Omega _{w}}}\sigma _{r_{1}r_{2}}^j) + H.c.}],
\end{align}
where the operator $L_j$ describes the decay of the system, and $\mathcal{L}$ is a Liouvillian superoperator. The Liouvillian superoperator can be regarded as an effective non-Hermitian operator corresponding to the density matrix, generating the non-unitary evolution of the density matrix. The non-Hermitian Hamiltonian of the system exhibits particle-hole symmetry. However, the closure of the $E_2$ and $E_3$ energy gaps leads to the breaking of particle-hole symmetry in the system, resulting in symmetric and symmetry-broken phases. Figure~\ref{FigA8}(e) is the phase diagram of the energy gap between $E_2$ and $E_3$ in the parameter plane of $\rm{\Omega_{w}}$ and $\rm{\Omega_{p}}$. The particle-hole symmetric region and the symmetry-broken region are separated by the gap-closing points (or exceptional points), in which the energy gap approaches zero along the white line. At the phase boundary line, the exceptional points exhibit a non-linear dependence on external parameters (e.g., electric field $\rm{\Omega_{w}}$ and probe field $\rm{\Omega_{p}}$). Furthermore, the Liouvillian eigenspectrum is obtained by solving $\mathcal{L}\hat{\rho}_i = \lambda_i \hat{\rho}_i$, where $\lambda_i$ and $\hat{\rho}_i$ denote the Liouvillian eigenvalues and eigenstates, respectively. The complex plane distribution of the Liouvillian eigenspectrum is shown in Fig.~\ref{FigA8}($\rm{f}_1$). As expected, the enlarged view in Fig.~\ref{FigA8}($\rm{f}_2$) demonstrates the emergence of an exceptional point resulting from the coalescence of eigenvalues and eigenstates.

\subsection{Exceptional point-enhanced Rydberg atomic electrometers}
\begin{figure}
	\centering
	\includegraphics[width=1\textwidth]{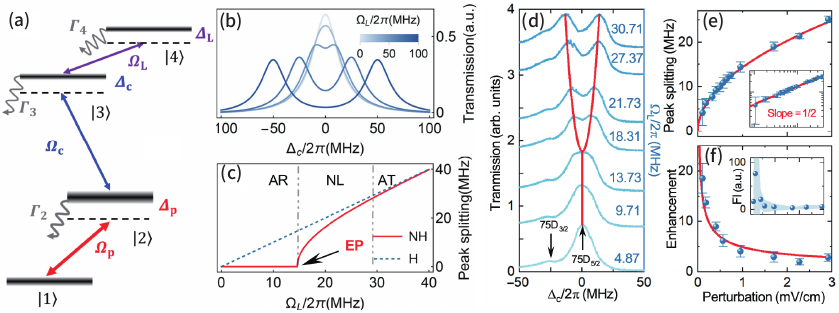}
	\caption{Observing exceptional points in Rydberg atomic ensembles (Figure adapted from Ref.~\cite{liang2026exceptional}). (a) Energy-level diagram of a four level Rydberg atomic system (b) Transmission spectra of the probe laser as a function of the coupling detuning $\rm{\Delta_c}$ for different values of the microwave Rabi frequency $\rm{\Omega_L}$. (c) Peak splitting as a function of the microwave Rabi frequency $\rm{\Omega_L}$. The blue dashed line corresponds to the Hermitian case and the red solid line represents the non-Hermitian case. The regions labeled AR, NL, and AT represent the absorption regime, nonlinear regime, and Autler-Townes regime, respectively. (d) Typical probe transmission spectra as a function of coupling detuning $\rm{\Delta_c}$ for increasing microwave Rabi frequency $\rm{\Omega_L}$ (from bottom to top). The red line traces the central peak positions, revealing nonlinear peak splitting behavior near the exceptional point. (e) Measured peak splitting versus perturbation strength $\rm{\Omega_s}$. (f) Measured enhancement factor as a function of perturbation strength.}
	\label{FigA9}
\end{figure}
Exceptional points in non-Hermitian systems enable approaches to ultrasensitive metrology. Chao Liang et al. introduced a theoretical framework for Rydberg electrometry based on the Autler-Townes effect under non-Hermitian conditions~\cite{liang2026exceptional}. They experimentally demonstrated a Rydberg atomic voltmeter that exploits exceptional points for sensitivity enhancement, based on a hot vapor cell. A second-order exceptional point was realized in an effective passive non-Hermitian Hamiltonian, achieved by harnessing the intrinsic dissipation of Rydberg excited states and optimizing the coupling optical fields. In this non-Hermitian system, the Autler-Townes splitting inherently exhibits nonlinear characteristics. Near the exceptional point region, the sensor's responsivity to weak microwave fields can be enhanced by nearly a factor of 20. By coupling the signal microwave field with a controllable local dressing field, amplitude- and phase-sensitive detection was realized. In the nonlinear Autler-Townes region, they observed amplified responses of the Autler-Townes splitting, achieving a sensitivity of 22.68 $\rm{nV~ cm^{-1}Hz^{-1/2}}$. This result proposes a paradigm for enhanced quantum electrometry by exceptional points, combining the intrinsic field sensitivity of Rydberg atoms with non-Hermitian criticality.

The model is based on the general four-level scheme for Rydberg electrometry. Rydberg atoms are excited using a two-photon excitation scheme, as shown in Fig.~\ref{FigA9}(a). The effective non-Hermitian Hamiltonian of the system reads~\cite{liang2026exceptional}:
\begin{align}
	\mathcal{H}_{\mathrm{NH}}=\frac{1}{2}\left(\begin{array}{ccc}
		-i \gamma_{\mathrm{p}} & -i \Omega_{\mathrm{eff}} & 0 \\
		i \Omega_{\mathrm{eff}} & -i \gamma_{\mathrm{c}} & -\Omega_{\mathrm{L}} \\
		0 & -\Omega_{\mathrm{L}} & 0
	\end{array}\right).
\end{align}
The emergence of the exceptional point divides the system into parity-time symmetry and breaking phases. Far from the exceptional point, the splitting becomes linearly proportional to the perturbation field $\Delta f={\rm{Re}}(E_+-E_-)\propto \rm{\Omega_s}$. Near the exceptional point, the system exhibits a nonlinear response to the electric field, characterized by a square-root scaling of the energy level splitting $\Delta f={\rm{Re}}(E_+-E_-)\propto \sqrt{\rm{\Omega_s}}$. Figure \ref{FigA9}(b) is a theoretical simulation of the traditional Autler-Townes splitting in a Hermitian system. The peak-to-peak splitting as a function of $\rm{\Omega_L}$ is plotted in Fig.~\ref{FigA9}(c) (red curve), with the Hermitian case indicated by the blue dashed line. Three distinct regimes are identified: an absorption regime in the parity-time broken phase; a nonlinear regime near the exceptional point in the parity-time symmetric phase; and a conventional Autler-Townes regime farther away from the exceptional point. Representative experimental results are shown in Fig.~\ref{FigA9}(d), where the transmission spectra of the probe field are measured as a function of the detuning $\rm{\Delta_c}$. From bottom to top, the strength of the microwave field $\rm{\Omega_L}$ gradually increases, leading to the progressive splitting of the resonance peak. The red line traces the central peak positions, revealing a nonlinear peak splitting behavior. Figure~\ref{FigA9}(e) demonstrates a square-root dependence of the peak splitting on the perturbation electric field near the exceptional point. The experimental results exhibit an enhancement of nearly 20-fold in responsivity in the vicinity of the exceptional point, as shown in Fig.~\ref{FigA9}(f).
\section{Topological states and non-Hermitian quantum phase transitions}\label{sec4}
Topological phases reveal the intrinsic spatial properties preserved by matter during continuous deformations such as stretching and bending, with transitions characterized by changes in topological invariants~\cite{kane2005z,konig2007quantum}. The rich properties of Rydberg atoms enable them to serve as a versatile platform for exploring topological states and phase transitions, while simultaneously introducing greater complexity to the system. Currently, a major challenge is to achieve these phenomena in interacting many-body platforms. Investigating the interaction mechanisms is therefore crucial for understanding how strong interactions and dissipation jointly govern topological phases in open quantum systems. 

\subsection{Observation of non-Hermitian topology in cold Rydberg quantum gases}
Recently, Jun Zhang et al. experimentally demonstrated non-Hermitian spectral topology in dissipative cold Rydberg atomic gases~\cite{zhang2025observation}. To characterize the spectral topology of the system, they presented the Liouvillian space $\mathcal{L}(\rm{Z})$ depends on the complex parameter $\rm{Z}=\rm{\Delta_r}+\mathit{i}\rm{\Omega_p^2}$. The results of diagonalizing the system's Liouvillian matrix are shown in Fig.~\ref{FigA10}(a), where the trajectory of eigenvalues on the complex plane reveals a `point-gap' topology. As $\rm{Z}$ varies along a closed contour $C_\mathrm{Z}$ in the parameter space, the Liouvillian eigenvalue forms a loop in the complex plane. This loop can wind around a reference energy point $\lambda$, giving rise to a nontrivial spectral winding number $\mathcal{W}_\lambda$, which is defined as \cite{zhang2025observation,budich2020non,wang2021generating}: 

\begin{figure}
	\centering
	\includegraphics[width=1\textwidth]{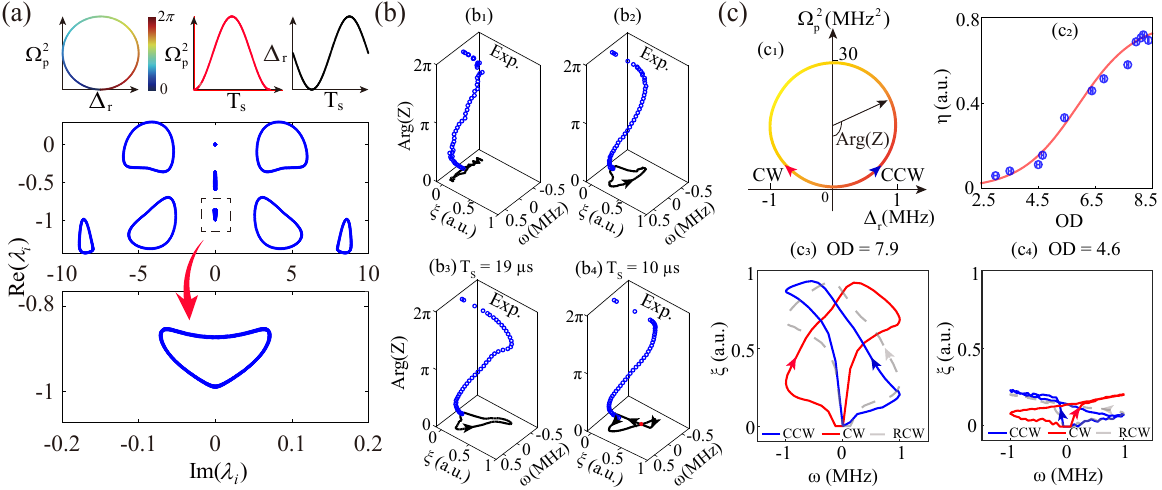}
	\caption{Non-Hermitian spectral topology (Figure adapted from Ref.~\cite{zhang2025observation}). (a) Numerical simulations of Liouvillian eigenspectrum varying with parameter $\rm{Arg(Z)}$ on the complex energy plane, where $\text{Z}={\rm{\Delta_r}}+i\rm{\Omega_{p}^2}$ and ${\rm{\Delta_r}}={\rm{\Delta_p}}+{\rm{\Delta_c}}$. The enlarged view represents the eigenvalue  $\lambda_{rr}$ of eigenstate $\rho_{rr}$. The upper panel shows the diagram of changing trends of parameters $\rm{\Omega^2 _p}$ and $\rm{\Delta _r}$. (b) The evolution trajectory of Rydberg eigenstate in the complex energy plane is strongly correlated with the variation of probe transmission and the energy level detuning $\rm{\Delta_r}$ in the experiment varies with the parameter Arg(Z), where $\text{Z}=\rm{\Delta_r}+\mathit{i}\rm{\Omega^2_{p}}$ in the scanning time of $\rm{T_s}=18~\rm{\mu s}$. Panels ($\rm{b_1}$), ($\rm{b_2}$), display the spectral features for $(\rm{\Omega_p}/2\pi)^2\in[0,1]~\rm{MHz}^2$, and $(\rm{\Omega_p}/2\pi)^2\in[0,4]~\rm{MHz}^2$. The scanning times are as follows: $\rm{T_s}=19~\rm{\mu s}$ in ($\rm{b_3}$), and $\rm{T_s}=10~\rm{\mu s}$ in ($\rm{b_4}$). (c) Chirality symmetry breaking. ($\rm{c_1}$) The clockwise (CW) and counterclockwise (CCW) scan paths of the parameters $\rm{\Omega_p^2}$ and $\rm{\Delta_r}$, corresponding to the chiral operation. ($\rm{c_2}$) The nonreciprocity parameter $\eta$ as a function of OD, and the red fitting curve. ($\rm{c_3}$, $\rm{c_4}$) The projections of spectra under two different scanning paths on the parameter plane, at the condition of optical depth OD = 7.9 and OD = 4.6. The grey dash curve denotes the mirror curve of the projected topological trajectory under the clockwise scanning.}
	\label{FigA10}
\end{figure}
\begin{align}
	\mathcal{W}_\lambda=\frac{1}{2 \pi i} \oint_{C_\mathrm{Z}} \mathrm{~d} \vec{\mathrm{Z}} \cdot \nabla_\mathrm{Z} \ln \operatorname{det}\left(\mathcal{L}(\mathrm{Z})-\mathrm{\lambda I}\right).
\end{align}
It counts the number of times the spectrum winds around the reference eigenvalues $\lambda$ and I is an identity matrix of the same dimension as $\mathcal{L}$. For a clockwise winding, $\mathcal{W}_\lambda = -1$, while for a counterclockwise winding, $\mathcal{W}_\lambda = 1$.

Experimentally, the topological evolution of the energy spectrum in the complex plane reveals a transition from a Hermitian state (Fig.~\ref{FigA10}($\rm{b_1}$)) to a non-Hermitian state (Fig.~\ref{FigA10}($\rm{b_2}$)). Notably, the scanning time governs the dynamic evolution of this topological structure by allowing for the tuning of the effective Rydberg atomic interactions. By tuning the scanning time, the spectral loop can be distorted, leading to a topological phase transition accompanied by a change in the sign of the winding number (see Figs.~\ref{FigA10}($\rm{b_3}$) and ($\rm{b_4}$)). Additionally, by reversing the direction of parameter scanning, differences in spectral loops can be observed (Fig.~\ref{FigA10}(c)). The distinct separation of the two curves highlights the asymmetry and path-dependence of the system’s response, revealing the characteristics of chiral symmetry breaking. The nonreciprocal parameter is defined as $\eta=\oint_{C_\mathrm{Z}} \mathrm{d} \vec{\mathrm{Z}}~{{\left| \mathit{O}_{\mathrm{RCW}}(\rm{Z}) - \mathit{O}_{\mathrm{CCW}}(\rm{Z})\right|}}$ to quantify the property of the chirality symmetry breaking of the system. Here, $O_{\mathrm{RCW}}(\rm{Z})$ and $O_{\mathrm{CCW}}(\rm{Z})$ denote the measured observables along the mirror-symmetric clockwise (RCW) and counterclockwise scanning paths at the complex parameter Z. The parameter $\eta$ quantifies the relative asymmetry between the system’s responses under these two scanning directions. Figure~\ref{FigA10}($\rm{c_2}$) directly presents the variation of the nonreciprocal parameter $\eta$ with OD.

\begin{figure}
	\centering
	\includegraphics[width=1\textwidth]{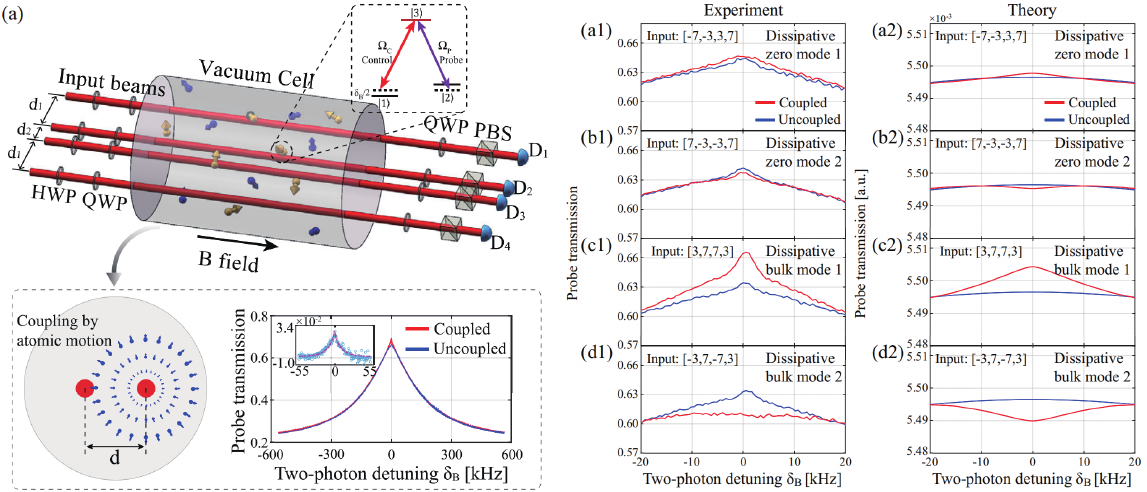}
	\caption{Schematics and principle of atomic vapor cell experiment simulating Su-Schrieffer-Heeger model with dissipative couplings (Figure adapted from Ref.~\cite{hao2023topological}). (a) Setup. Top: several optical channels with designed spacings in the vaporcell create ground state coherences (spin waves) by electromagnetically induced transparency process in each channel. Bottom right: characterization of dissipative coupling rate using a two-channel setting. (a1)-(d1) Measured electromagnetically induced transparency spectra via the probe transmission (normalized to far off resonant 100\% transmission), coupled (all channel probes on) and uncoupled (only probe in the detected channel on), for the four input states [$-7$, $-3$, $3$, $7$]$^T$, [$7$, $-3$, $-3$, $7$]$^T$, [$3$, $7$, $7$, $3$]$^T$, [$-3$, $7$,$-7$, $3$]$^T$, respectively. (a2)-(d2) Theoretical calculations of the optical coherences.}
	\label{FigA11}
\end{figure}

\subsection{Topological atomic spin wave lattices by dissipative couplings}
Topological phases of quantum matter also exhibit fascinating phenomena, such as edge modes unaffected by defects, which hold potential applications in quantum computing~\cite{nayak2008non,field2018introduction}. The robustness of edge modes across various lattice systems can largely be attributed to energy gap protection and their connection to topologically nontrivial bands. Symmetry-protected edge modes remain energetically located within the bulk energy gap and are immune to local perturbations. Recently, lattice systems with dissipative coupling exhibit spectral characteristics distinct from coherently coupled networks and may enable topological dissipation. Therefore, dissipative coupling also offers the possibility of designing topological structures.

Generally, the Su-Schrieffer-Heeger model is a Hermitian model with a one-dimensional dimerized chain of even size: each unit cell contains two lattice sites (labeled A and B), with intracell coupling strength denoted as $J_1$ and intercell coupling strength denoted as $J_2$. In the tight-binding approximation, the Hamiltonian is given by 
\begin{equation}
	\begin{aligned}
		H=\sum_n(J_1|n, B\rangle\langle n, A|+J_2|n+1, A\rangle\langle n, B|+H.c.). 
	\end{aligned}
\end{equation}
The topological properties of this model are entirely determined by the ratio $J_1/J_2$. When $J_1<J_2$, the system is a topologically nontrivial phase with a winding number of $1$, supporting topologically protected zero-energy edge states at both ends of the chain. When $J_1>J_2$, the system is in a topologically trivial phase with no edge states. When $J_1=J_2$, the band gap closes, marking the topological phase transition point.

However, Dong-Dong Hao et al. experimentally realized a dissipative version of the Su-Schrieffer-Heeger model~\cite{hao2023topological}. Based on a spatial lattice of atomic spin waves in a vapor cell, the experiment utilized an enriched $^{87}\rm{Rb}$ vapor cell [Fig. \ref{FigA11}(a)]. The cell temperature is set to 40$^\circ$C to maintain a relatively small optical depth. The laser beam was split into several spatially separated beams via polarization-maintaining fibers, forming optical channels within the vapor cell. Each channel established a $\Lambda$-type electromagnetically induced transparency configuration involving ground states $\ket{1}$, $\ket{2}$, and Rydberg state $\ket{3}$. A uniform magnetic field was applied to induce Zeeman shifts in the energy levels, enabling precise tuning of the two-photon detuning $\rm{\delta_B}$ (the illustration in Fig. \ref{FigA11}(a)). Spin waves in different optical channels were coupled inhomogeneously through atomic motion. Consequently, a dissipative version of the Su-Schrieffer-Heeger model was implemented by arranging channels with alternating spacings $d_1$ and $d_2$. The non-Hermitian Hamiltonian can be written as~\cite{hao2023topological}:
\begin{equation}
	\begin{aligned}
		H = &i v e^{i \delta_B d_1 / \nu} \sum_m \left( |a, m\rangle \langle b, m| + |b, m\rangle \langle a, m| \right) \\
		&+i w e^{i \delta_B d_2 / \nu} \sum_m \left( |b, m\rangle \langle a, m+1| + |a, m+1\rangle \langle b, m| \right),
	\end{aligned}
\end{equation}
where the dissipative coupling rates within and between unit cells, $v$ and $w$, satisfy  $v/w \propto d_2/d_1$. When $\delta_B=0$, a purely dissipative version of the typical Su-Schrieffer-Heeger model is realized, which possesses chiral symmetry and exhibits a topological dissipative spectrum when $v< w$. This is characterized by edge modes existing within the bulk dissipative gap at zero dissipation rate, whereas the case where $v>w$ is trivial.

In the experiment, transmission spectra for both channels were measured at a relatively high probe power to improve the signal-to-noise ratio, thereby allowing for the extraction of the average eigenvalue. The results show that the electromagnetically induced transparency spectrum for the edge state input nearly coincides with the uncoupled electromagnetically induced transparency spectrum, as shown in Figs. \ref{FigA11}(a1)-(b1). In contrast, for the bulk modes in Figs. \ref{FigA11}(c1)-(d1), significant changes in both peak intensity and linewidth of the electromagnetically induced transparency spectrum are observed compared to the uncoupled case. Figures \ref{FigA11}(a2)-(d2) show the theoretical results obtained by calculating the Hamiltonian, which are consistent with the experimental findings.

\begin{figure}
	\centering
	\includegraphics[width=0.9\textwidth]{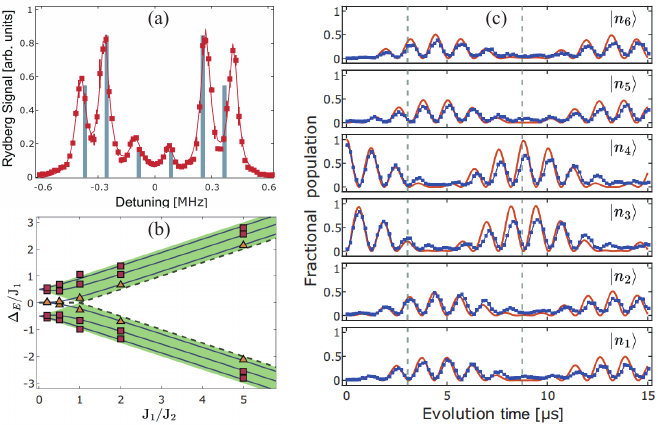}
	\caption{Probing the topological phase transition using Rydberg atom synthetic dimensions (Figure adapted from Ref.~\cite{lu2024probing}). (a) Rydberg excitation spectrum with the laser frequency scanned across the bare state $\ket{n_2}$ for tunneling ratio $J_1/J_2 =1$. (b) The eigenenergies measured from the peak positions $\Delta_{E}$ in the Lorentzian fits of the Rydberg excitation spectra at different tunneling ratios $J_1/J_2 =1$. The triangles and squares denote the edge (at small J1/J2) and bulk states, respectively, and the solid lines show the calculated results. (c) Evolution of the fractional populations in each $n^3 S_1$ bare Rydberg state with initial excitation to the fourth state $\ket{n_4}$.}
	\label{FigA12}
\end{figure}
\subsection{Probing the topological phase transition using Rydberg atom synthetic dimensions}
A synthetic dimension refers to the use of internal degrees of freedom of a quantum system (such as Rydberg energy levels of atoms, spin states, momentum states, etc.) to simulate spatial dimensions in real space~\cite{ozawa2019topological,arguello2024synthetic}. In Rydberg atomic systems, Rydberg states with high principal quantum numbers possess dense energy level structures, and these energy levels can serve as lattice sites in a synthetic lattice. Synthetic dimensions based on Rydberg atoms offer another highly flexible experimental platform for realizing and probing topological phase transitions. Y. Lu et al. experimentally investigated the topological phase transition in the Su-Schrieffer-Heeger Hamiltonian by constructing a six state synthetic dimension from Rydberg atomic levels~\cite{lu2024probing}. This work demonstrates that even with a finite-sized system, the Su-Schrieffer-Heeger model implemented via Rydberg atom synthetic dimensions enables reliable observation of topological phase transitions through combined spectral and dynamical measurements. The platform is highly extensible: larger systems can be realized by incorporating more Rydberg levels, and more complex lattice geometries or artificial gauge fields can be engineered. This approach opens the experimental avenue for studying topological states and their dynamics in Rydberg atom synthetic dimensions.

In this experiment, a cold gas of $^{84}\mathrm{Sr}$ atoms was used, with six neighboring $n^3S_1$ Rydberg states ($n \approx 60$) serving as sites in a synthetic lattice. Adjacent states were coupled via two photon microwave transitions, implementing a Su-Schrieffer-Heeger model with alternating strong ($J_2$) and weak ($J_1$) tunneling amplitudes. By tuning the microwave field amplitudes, the tunneling ratio \(J_1/J_2\) could be precisely controlled, enabling systematic exploration of the topological phase transition across the range $J_1/J_2 = \{0.2, 0.5,1,2,5\}$.

By measuring the Rydberg excitation spectrum in the presence of microwave fields, the evolution of the eigenenergy spectrum with the tunneling ratio was directly observed (Figs.~\ref{FigA12}(a)-(b)). At small values of $J_1/J_2$, the spectrum exhibits nearly degenerate zero-energy edge states. As $J_1/J_2$ increases and crosses the critical value, these edge states disappear, and a bulk band gap opens, further confirming the loss of topological protection. Despite the system consisting of only six states, it captures the essential features of the Su-Schrieffer-Heeger model, including the bulk band structure, edge states, and their behavior near the phase transition. To probe the topological properties of the system, the team combined quench dynamics experiments with spectral measurements. In the quench experiments, the atom was initially excited to a selected Rydberg state, after which the microwave couplings were switched on, and the time evolution of the population in each state was recorded (Fig.~\ref{FigA12}(c)). 

\subsection{Observation of non-Hermitian many-body phase transition in a Rydberg-atom array}
\begin{figure}
	\centering
	\includegraphics[width=1\textwidth]{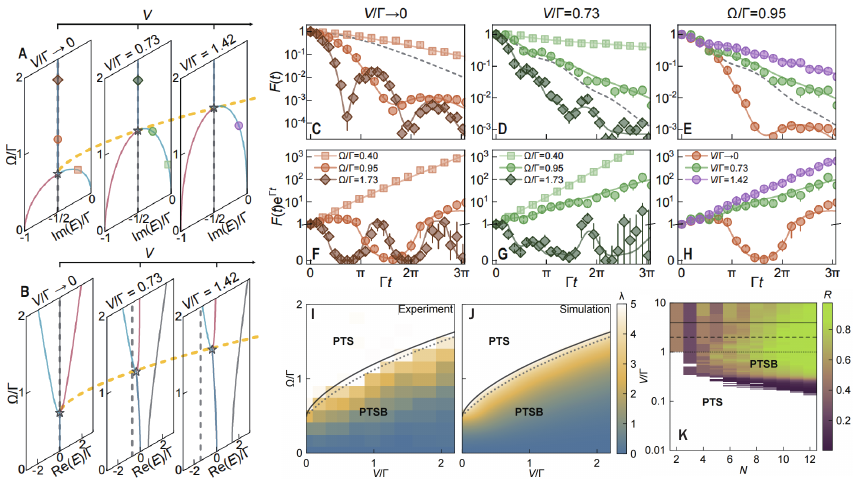}
	\caption{Non-Hermitian spectra, dynamics of Loschmidt Echo, and parity-time phase diagram of interacting Rydberg atoms. (Figure adapted from Ref.~\cite{zhang2025observation1}). (A-B) Imaginary (A) and real (B) parts of the eigenvalues $E$. (C-E) Evolution of the Loschmidt Echo (LE) $F(t)$. Solid curves are numerical simulations using the experimental parameters, while dashed curves are simulated results at the exceptional points. (F-H) Scaled Loschmidt Echo, $F(t){\rm{e}}^{\Gamma t}$, corresponding to the data in (C-E). (I-J) Parity-time phase diagram of interacting Rydberg atoms. Measured (I) and simulated (J) phase diagram of parity-time symmetry (PTS) and parity-time symmetry breaking (PTSB). (K) Parity-time phase transition of $N$ Rydberg atoms.}
	\label{FigA13}
\end{figure}

Rydberg atom arrays, leveraging their advantages of single-site addressability, highly tunable interactions, and flexible scalability in system size, have emerged as a suitable platform for controllable many-body quantum simulation. Yao-Wen Zhang et al. experimentally implemented a non-Hermitian $XY$ spin chain model in a one-dimensional programmable optical tweezer array~\cite{zhang2025observation1}, a system that exhibits the complex interplay of dissipation and coherent interactions. They observed the parity-time symmetry breaking phase transition, thereby revealing how interactions fundamentally shape non-Hermitian dynamics in a controllable atomic array. It establishes an experimental foundation for exploring non-Hermitian many-body physics beyond the single-particle paradigm in highly controllable quantum simulators.

The core of the experiment lies in engineering an effective non-Hermitian Hamiltonian with parity-time symmetry through microwave field coupling and state-selective dissipation~\cite{zhang2025observation1}:
\begin{equation}
	H_{\mathrm{pt}} = \sum_{i=1}^{N} \left( \frac{\Omega}{2} \sigma_i^x + i\frac{\Gamma}{4} \sigma_i^z \right) + \sum_{i<j} \frac{V_{ij}}{2} \left( \sigma_i^x \sigma_j^x + \sigma_i^y \sigma_j^y \right),
\end{equation}
where $\sigma_i^{x,y,z}$ are Pauli operators of the $i$-th atom, $\Gamma$ is a state-selective dissipation generated by coupling two different Rydberg states, and $V_{ij} \propto r_{ij}^{-3}$ is the dipole-dipole exchange interaction. The numerical calculation of eigenvalues is shown in Figs.~\ref{FigA13}A and B. By precisely tuning the interatomic distance and the microwave Rabi frequency $\Omega$, the system can undergo a phase transition between the parity-time symmetric and symmetry breaking phases. To probe this transition, the research team introduced the Loschmidt Echo as a dynamical observable, defined as the survival probability of the initial state $|0_N\rangle = |\uparrow\uparrow\cdots\uparrow\rangle$ after non-Hermitian evolution: $F(t) = |\langle 0_N | \psi_N(t) \rangle|^2$. Experimentally, the measured Loschmidt Echo displays damped oscillations in the parity-time symmetry phase, whereas it follows monotonic exponential decay in the symmetry breaking phase, reflecting the transition from oscillatory to dissipative dynamics (Figs.~\ref{FigA13}C-E). The scaled Loschmidt Echo $F(t)e^{\Gamma t}$ more clearly reveals persistent oscillations in the parity-time symmetry phase and exponential gain in the symmetry breaking phase (Figs.~\ref{FigA13}F-H). 

Notably, the experiment also observed an interaction-driven shift of the exceptional point (Figs.~\ref{FigA13}I-J): as the interaction strength $V$ increases, the critical Rabi frequency $\Omega_c$ corresponding to the exceptional point rises significantly, demonstrating the ability of many-body interactions to tune the non-Hermitian phase boundary. Further investigation of the dynamical behavior in multi-atom chains ($N \geq 3$) can reveal richer many-body effects, such as size-dependent non-Hermitian blockade. Researchers also prepared a linear array of $N$ Rydberg atoms to experimentally study the size-dependent effects in the parity-time symmetry and breaking phases, as shown in Fig.~\ref{FigA13}K.

\section{Summary}\label{sec5}
This review mainly summarizes recent progress in both non-Hermitian physics and Rydberg atomic systems. Non-Hermitian systems~\cite{yokomizo2019non,ashida2020non,helbig2020generalized,li2025realization,wang2026localization}, originating from gain, loss, and nonreciprocal coupling, display a range of distinctive phenomena: complex energy spectra, a biorthogonal basis~\cite{zhang2022review}, symmetric and symmetry-broken phases~\cite{kawabata2019symmetry}, and nontrivial topological phases~\cite{wang2021topological}. Representative phenomena include spontaneous symmetry breaking, the emergence of exceptional points~\cite{bergholtz2021exceptional,zhou2018observation}, sensing enhanced by exceptional points, and non-Hermitian topological structures. These developments have been widely demonstrated in optical~\cite{barik2018topological,parto2020non}, acoustic~\cite{huang2024acoustic}, electrical~\cite{zhang2020non}, and cold atom platforms~\cite{li2020topological,yoshida2018reduction}.

Rydberg atoms, with their strong interactions and high controllability, have become an important platform for exploring non-Hermitian quantum many-body physics. Many-body interactions can induce effective dissipation, giving rise to tunable non-Hermitian Hamiltonians. The experimental observation of interaction-induced exceptional points, charge-conjugate parity symmetry breaking, and dynamical features such as non-Hermitian hysteresis trajectories have been reported~\cite{zhang2025exceptional}. Additionally, the role of Liouvillian exceptional points in open quantum dynamics and their connection to chiral state transfer has been demonstrated~\cite{xie2025chiral}. Moreover, microwave-induced trajectory intersections provide a mechanism for highly sensitive microwave detection~\cite{wang2026quantum}. The implementation of Rydberg electrometers leveraging exceptional points has demonstrated a substantial enhancement in electric field sensing sensitivity~\cite{liang2026exceptional}. 

Recent experiments on dissipative cold Rydberg gases have demonstrated non-Hermitian spectral topology, characterized by non-trivial winding of the eigenvalue spectrum in the complex energy plane~\cite{zhang2025observation}. Furthermore, by exploiting motion-induced dissipative couplings, researchers have realized a dissipative topological structure (Su-Schrieffer-Heeger model), enabling the observation of topological edge states~\cite{hao2023topological}. Meanwhile, Rydberg atom synthetic dimensions provide an alternative pathway to realize and probe the Su-Schrieffer-Heeger model, where phase transitions and topological invariants can be reliably extracted even in finite-size systems through combined spectral and dynamical measurements~\cite{lu2024probing}. The programmable atom array platform has demonstrated the observation of non-Hermitian many-body phase transitions and interaction-driven exceptional point shifts, revealing phenomena such as Loschmidt echo dynamics and size-dependent non-Hermitian many-body blockade~\cite{zhang2025observation1}. These studies collectively establish Rydberg atoms as a powerful platform for investigating topological phenomena, non-Hermitian phase transitions, and critical dynamics in open quantum systems.

Currently, significant progress has been made in the study of non-Hermitian physics based on Rydberg atomic systems, yet this field remains in its early stages of rapid development. Future research can further focus on Liouvillian spectral engineering, quantum state and non-Abelian manipulation~\cite{gao2025photonic,wang2026observation}. Compared with the effective non-Hermitian Hamiltonian, the full Liouvillian dynamics can provide a more comprehensive characterization of quantum jump processes, metastable structures, and steady-state phase behaviors. Furthermore, the highly programmable platform based on Rydberg atom arrays and synthetic dimensions can be used to construct tunable non-Hermitian lattice models and higher-order exceptional point structures~\cite{wang2023experimental,yan2025ultrasensitive}, offering key technical support for scalable quantum simulation of open quantum systems. On the application front, leveraging the high sensitivity response near exceptional points, non-Hermitian Rydberg systems show significant potential for quantum precision measurement and sensing. Future research can further enhance their noise resilience, optimize parameter estimation strategies, and extend to multi-parameter joint measurements. In addition, combining the strong nonlinearity of Rydberg atoms with non-Hermitian effects may enable the development of novel quantum information processing schemes.
	
	\section*{Acknowledgements}
	We acknowledge funding from the National Key R and D Program of China (Grant No. 2022YFA1404002), the National Natural Science Foundation of China (Grant Nos. T2495253, 62435018, 125B2085)
	
	%\funding{Sample text inserted for demonstration.}
	% This section is a list of funder names and grant numbers
	
	\section*{Author contributions statement}
	This review was written by Ya-Jun Wang and Jun Zhang under the guidance of Dong-Sheng Ding. All authors contributed to discussions regarding the results and the analysis contained in the manuscript.

\end{document}